\documentclass{aa}

\usepackage{graphicx}
\usepackage{CJK}
\usepackage{enumitem}
\usepackage[varg]{txfonts}
\usepackage{amsmath}
\usepackage{amssymb}
\usepackage{footmisc}
\usepackage{booktabs}
\usepackage[colorlinks = true,
            linkcolor = blue,
            urlcolor  = blue,
            citecolor = blue,
            anchorcolor = blue,
            unicode]{hyperref}

\begin{document}

\begin{CJK*}{UTF8}{gbsn}

\title{Evolved Massive Stars at Low-metallicity \uppercase\expandafter{\romannumeral2}.\\ Red Supergiant Stars in the Small Magellanic Cloud}
\titlerunning{Evolved Massive Stars at Low-Z \uppercase\expandafter{\romannumeral2}. RSGs in the SMC}

\author{
Ming Yang (杨明) \inst{1} \and Alceste Z. Bonanos \inst{1} \and Bi-Wei Jiang (姜碧沩) \inst{2} \and Jian Gao (高健) \inst{2} \and Panagiotis Gavras \inst{3} \and Grigoris Maravelias \inst{1} \and Shu Wang (王舒) \inst{4} \and Xiao-Dian Chen (陈孝钿) \inst{4} \and Frank Tramper \inst{1} \and Yi Ren (任逸) \inst{2} \and Zoi T. Spetsieri \inst{1} \and Meng-Yao Xue (薛梦瑶) \inst{5} 
}
\authorrunning{Yang, Bonanos \& Jiang et al.}

\institute{
IAASARS, National Observatory of Athens, Vas. Pavlou and I. Metaxa, Penteli 15236, Greece\\
                \email{myang@noa.gr} \and
Department of Astronomy, Beijing Normal University, Beijing 100875, People's Republic of China \and
Rhea Group for ESA/ESAC, Camino bajo del Castillo, s/n, Urbanizacion Villafranca del Castillo, Villanueva de la Cañada, 28692 Madrid, Spain \and
CAS Key Laboratory of Optical Astronomy, National Astronomical Observatories, Chinese Academy of Sciences, Datun Road 20A, Beijing 100101, People's Republic of China \and
International Centre for Radio Astronomy Research, Curtin University, Bentley, WA 6102, Australia
}

\abstract{
We present the most comprehensive red supergiant (RSG) sample for the Small Magellanic Cloud (SMC) up to now, including 1,239 RSG candidates. The initial sample is derived based on a source catalog for the SMC with conservative ranking. Additional spectroscopic RSGs are retrieved from the literature, as well as RSG candidates selected based on the inspection of \textit{Gaia} and 2MASS color-magnitude diagrams (CMDs). We estimate that there are in total $\sim$ 1,800 or more RSGs in the SMC. We purify the sample by studying the infrared CMDs and the variability of the objects, though there is still an ambiguity between asymptotic giant branch stars (AGBs) and RSGs at the red end of our sample. One heavily obscured target is identified based on multiple NIR and MIR CMDs. The investigation of color-color diagrams (CCDs) shows that, there are much less RSGs candidates ($\sim$4\%) showing PAH emission features compared to the Milky Way and LMC ($\sim$15\%). The MIR variability of RSG sample increases with luminosity. We separate the RSG sample into two subsamples (``risky'' and ``safe'') and identify one M5e AGB star in the ``risky'' subsample, based on simultaneous inspection of variabilities, luminosities and colors. The degeneracy of mass-loss rate (MLR), variability and luminosity of RSG sample is discussed, indicating that most of the targets with large variability are also the bright ones with large MLR. Some targets show excessive dust emission, which may be related to previous episodic mass loss events. We also roughly estimate the total gas and dust budget produced by entire RSG population as $\rm \sim1.9^{+2.4}_{-1.1}\times10^{-6}~M_{\odot}/yr$ in the most conservative case, according to the derived MLR from $\rm IRAC1-IRAC4$ color. Based on the MIST models, we derive a linear relation between $T_{\rm eff}$ and observed $\rm J-K_S$ color with reddening correction for the RSG sample. By using a constant bolometric correction and this relation, the Geneva evolutionary model is compared with our RSG sample, showing a good agreement and a lower initial mass limit of $\sim$7 $\rm M_\sun$ for the RSG population. Finally, we compare the RSG sample in the SMC and the LMC. Despite the incompleteness of LMC sample in the faint end, the result indicates that the LMC sample always shows redder color (except for the $\rm IRAC1-IRAC2$ and $\rm WISE1-WISE2$ colors due to CO absorption) and larger variability than the SMC sample, which is likely due to a positive relation between MLR/variability and the metallicity. 
}


\maketitle

\section{Introduction}

Red supergiant stars (RSGs) are moderately massive stars ($\sim8-30~M_\sun$) in the helium-burning evolutionary phase and located in the upper right (cool and luminous) region of the Hertzsprung-Russell (H-R) diagram. They are the result of core hydrogen exhausted main-sequence stars streak across the top of the H-R diagram. As a unique class of the massive star population, their young ages ($\rm \sim8-20~Myr$), low effective temperature ($T_{\rm eff} \sim3500-4500~K$), high luminosities ($\rm \sim4000-400000~L_\sun$), and large radii ($\rm \sim100-1000R_\sun$) represent a critical extremity of stellar evolution \citep{Humphreys1979, Levesque2005, Ekstrom2013, Massey2013, Davies2017}. 

In general, RSGs are considered to have two destinies, depending on the initial mass, chemical composition, and more importantly, the mass-loss rate (MLR). Some of them may stay in the RSG stage and eventually explode as hydrogen-rich Type \uppercase\expandafter{\romannumeral2}-P supernovae (SN). The others may evolve backwards to the blue end of the H-R diagram and spend some short periods of time as yellow supergiant stars (YSGs), blue supergiant stars (BSGs) or Wolf-Rayet stars (WRs) before the final SN explosion \citep{Smartt2009, Humphreys2010, Ekstrom2012, Meynet2015, Davies2018}. In any case, the strong mass loss during the RSG phase may largely influence the ultimate fate of the RSGs, and significantly contribute to the dust content and chemical enrichment of young stellar populations. Specially, RSGs may be one of the main contributors for dust production in young galaxies at high redshift, where the metallicity is much lower than the local Universe and the potential dust producer of asymptotic giant branch stars (AGBs) are not yet evolved \citep{Massey2005, Levesque2010}. However, the major physical mechanisms (e.g., episodic mass loss, stellar winds, pulsation, convection, luminosity, metallicity, binarity, etc.), which dominate the mass loss of RSGs are still unclear \citep{Macgregor1992, Harper2001, Yoon2010, Mauron2011, Beasor2016}. 

To better understand the nature of the RSGs, it is crucial to build a representative sample of RSGs covering large ranges of both metallicities and luminosities. For the low-metallicity environments, there are two excellent examples in the local Universe, the Large and Small Magellanic Cloud (LMC and SMC; about half and one-fifth of the metallicity of the Milky Way, respectively; \citealt{Russell1992, Rolleston2002, Keller2006, Dobbie2014, DOnghia2016}). Due to their proximity, each individual star in the MCs can be resolved, which results in numerous studies for the massive stars in the past half century \citep{Feast1980, Barba1995, Massey2003, Evans2008, Neugent2010, Bouret2013, Kourniotis2014, Hainich2015, Castro2018}. However, one of the obstacles for studying massive stars in the MCs is the foreground contamination, e.g, the low-mass red dwarfs in the Milky Way appear to have similar brightness to the RSGs in the MCs, and the contamination is even worse for the low-luminosity RSGs. As a result of the contamination and also the efficiency of the observation, previous studies only focused on the bright end of RSGs population. Fortunately, this problem can be largely solved by utilizing the \textit{Gaia} Data Release 2 (DR2) \citep{Gaia2016, Gaia2018}, for which we have built a multiwavelength source catalog and identified three evolved massive star populations (BSGs, YSGs, and RSGs) in the SMC \citep{Yang2019a}.

In this paper, we present the analysis of the most comprehensive RSG sample in the SMC up to now. The sample selection and data analysis are presented in \textsection2 and \textsection3, respectively. The discussion is described in \textsection4. The summary is given in \textsection5.

\section{Sample selection}

The main sample of RSG candidates in the SMC was derived from a source catalog for the SMC \citep{Yang2019a}. We only briefly describe here the dataset and more details can be found in the original paper. The SMC source catalog is a clean, magnitude-limited (IRAC1 or WISE1 $\leq$ 15.0 mag; \citealt{Werner2004, Wright2010}) multiwavelength source catalog with 45,466 targets in total. It contains data in 50 different bands including 21 optical and 29 infrared (IR) bands, ranging from ultraviolet to far-IR. Additionally, radial velocities and spectral classifications are collected from the literature, as well as the IR and optical variability statistics are derived from different datasets. The catalog was essentially built upon a $1''$ crossmatching and a $3''$ deblending between the \textit{Spitzer} Enhanced Imaging Products (SEIP) source list and \textit{Gaia} DR2 photometric data. We removed the foreground contamination by further constraining the proper motions and parallaxes from \textit{Gaia} DR2. By using the evolutionary tracks and synthetic photometry from Modules for Experiments in Stellar Astrophysics (MESA; \citealt{Paxton2011, Paxton2013, Paxton2015, Paxton2018}) Isochrones \& Stellar Tracks (MIST\footnote{http://waps.cfa.harvard.edu/MIST/}; \citealt{Choi2016, Dotter2016}) and also the theoretical $\rm J-K_S$ color cuts \citep{Cioni2006a, Boyer2011}, we identified three evolved massive star populations of BSGs, YSGs, and RSGs in the SMC from five different color-magnitude diagrams (CMDs). There are 1,405 RSG, 217 YSG and 1,369 BSG candidates, respectively. We ranked the candidates based on the intersection of different CMDs, where Rank 0 was given to a target identified as the same type of evolved massive star in all five CMDs by the MIST models and so on, and Rank 5 indicated the additional RSG candidates identified by the theoretical $\rm J-K_S$ color cuts but not recovered by the MIST models. 

The main RSG sample contains 1,405 candidates from the SMC source catalog. However, targets with Rank 4 and 5 are selected only in one CMD by either the MIST models or the theoretical $\rm J-K_S$ color cuts, and many of them reach down close to the tip of the red giant branch (TRGB) and AGBs population (see also Figure 13 and 14 of \citealt{Yang2019a}). To be on the safe side, we adopted only targets with ranks from 0 to 3 (targets identified in at least two CMDs) as our initial sample, which resulted in 1,107 targets. Due to the photometric quality cuts and uncertainties of the \textit{Spitzer} and \textit{Gaia} data, the strict constraints on the astrometric solution, and the deblending applied during the construction of the source catalog, some of the spectroscopically confirmed RSGs were also rejected. In order to make the sample as complete as possible, we retrieved and added all known spectroscopic RSGs in both optical and mid-infrared (MIR) bands from Simbad \citep{Wenger2000} and data taken by \textit{Spitzer} Infrared Spectrograph (IRS; \citealt{Houck2004}), respectively. From Simbad, we selected 322 RSGs with RV $\geq$ 90 km/s, spectral type later than G0 and luminosity class brighter than \uppercase\expandafter{\romannumeral2} by using criteria query \citep{Levesque2013, Gonzalez2015}, for which 192 targets were matched with our initial sample within $1''$. Additionally, a crossmatching with the main RSG sample of 1,405 candidates indicated that three Rank 4 candidates were also matched within $1''$. Surprisingly, there are two spectroscopic RSGs matched with the source catalog within $1''$, but not selected as the RSG candidates by either the MIST models or the theoretical $\rm J-K_S$ color cuts. Visual inspection of \textit{Gaia} and 2MASS (Two Micron All Sky Survey; \citealt{Skrutskie2006}) CMDs (shown below) indicated that these two targets were slightly off the blue and red boundaries of the RSG region, respectively, which was likely due to the intrinsic variability of the RSGs \citep{Kiss2006, Yang2011, Ren2019}. Consequently, in total, there are 127 unselected spectroscopic RSGs from Simbad. For data taken by \textit{Spitzer}/IRS, there were 22 RSGs from \citet{Ruffle2015}, who classified 209 point sources observed by \textit{Spitzer}/IRS using a decision tree method, based on IR spectral features, continuum and spectral energy distribution shape, bolometric luminosity, cluster membership and variability information (all the targets from \citealt{Kraemer2017} were also included). Among those 22 RSGs, 16 of them were matched with our initial sample within $1''$, and four of them were matched with the previous unselected Simbad RSGs within $1''$. Thus, there are only two unselected spectroscopic RSGs from \textit{Spitzer}/IRS. In total, there are additional 129 spectroscopic RSGs from both Simbad and \textit{Spitzer}/IRS, for which we give them Rank -1. 

Since the additional crossmatching between Simbad and the main RSG sample indicated that three Rank 4 candidates were also spectroscopic RSGs, it occurred to us that there might be more RSGs among the Rank 4 candidates. However, due to lack of spectroscopic data, we are only able to identify RSG candidates with extreme red colors and high luminosities. This is due to the fact that, there are probably only two kinds of candidates in the luminous red end, RSGs and super-AGBs (stars in the mass range of $\sim7-10~M_{\sun}$ that represent a transition to the more massive supergiant stars and are characterized by degenerate off-centre carbon ignition analogous to the earlier helium flash; \citealt{Herwig2005, Siess2006, Groenewegen2009, Doherty2017}). We also expect that super-AGBs could be rejected by using several methods as shown in \citet{Yang2018}. As a result, another three Rank 4 CMD candidates were identified by simultaneous visual inspection of \textit{Gaia} and 2MASS CMDs (shown below), for which they were all brighter than the AGBs branch in both CMDs, except one in the 2MASS CMD. However, the exceptional target is located well inside the RSGs region, even though it has similar luminosity compared to the bright end of AGBs population. 

\begin{figure*}
\center
\includegraphics[bb=54 425 558 635]{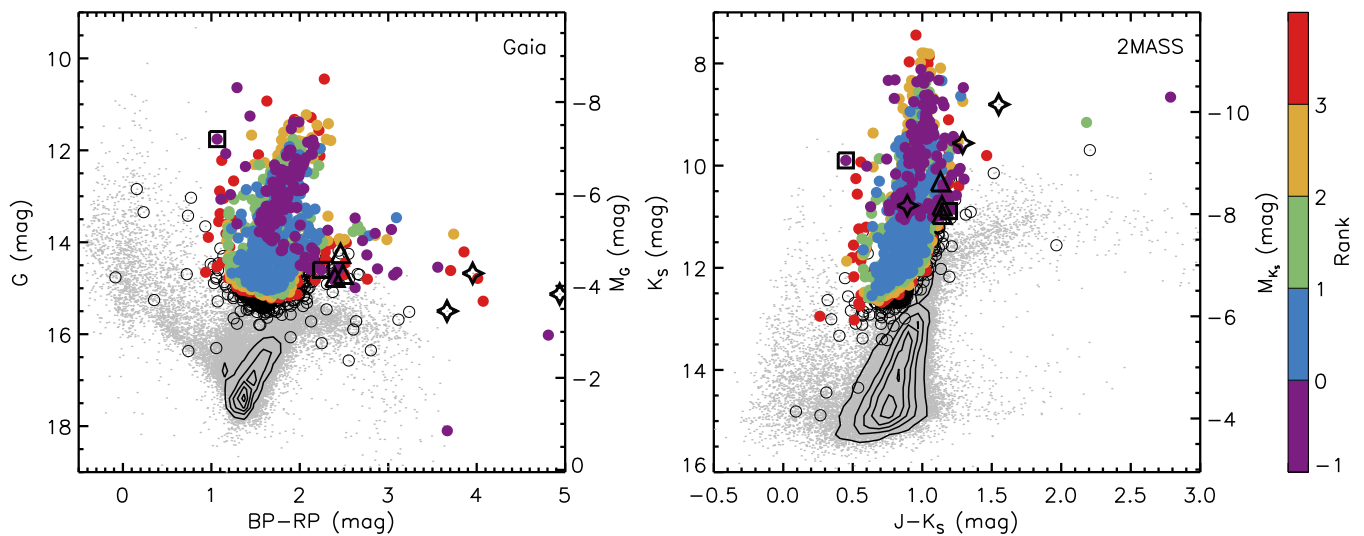}
\caption[]{Color-magnitude diagrams of \textit{Gaia} and 2MASS datasets. Targets from the SMC source catalog are shown as gray dots. The valid RSG candidates are shown as solid circles and color coded ranging from Rank -1 to 3. Three Rank 4 spectroscopic RSGs are shown as open triangles, while the other three Rank 4 CMD RSG candidates are shown as open stars. Two unselected spectroscopic RSGs are shown as open squares. The rest of Rank 4 and 5 targets from the main RSG sample are shown as open circles. Black contours represent the number density of the SMC source catalog. See text for details. \label{cmd_gaia_tmass}}
\end{figure*}

\begin{figure}
\includegraphics[scale=0.3]{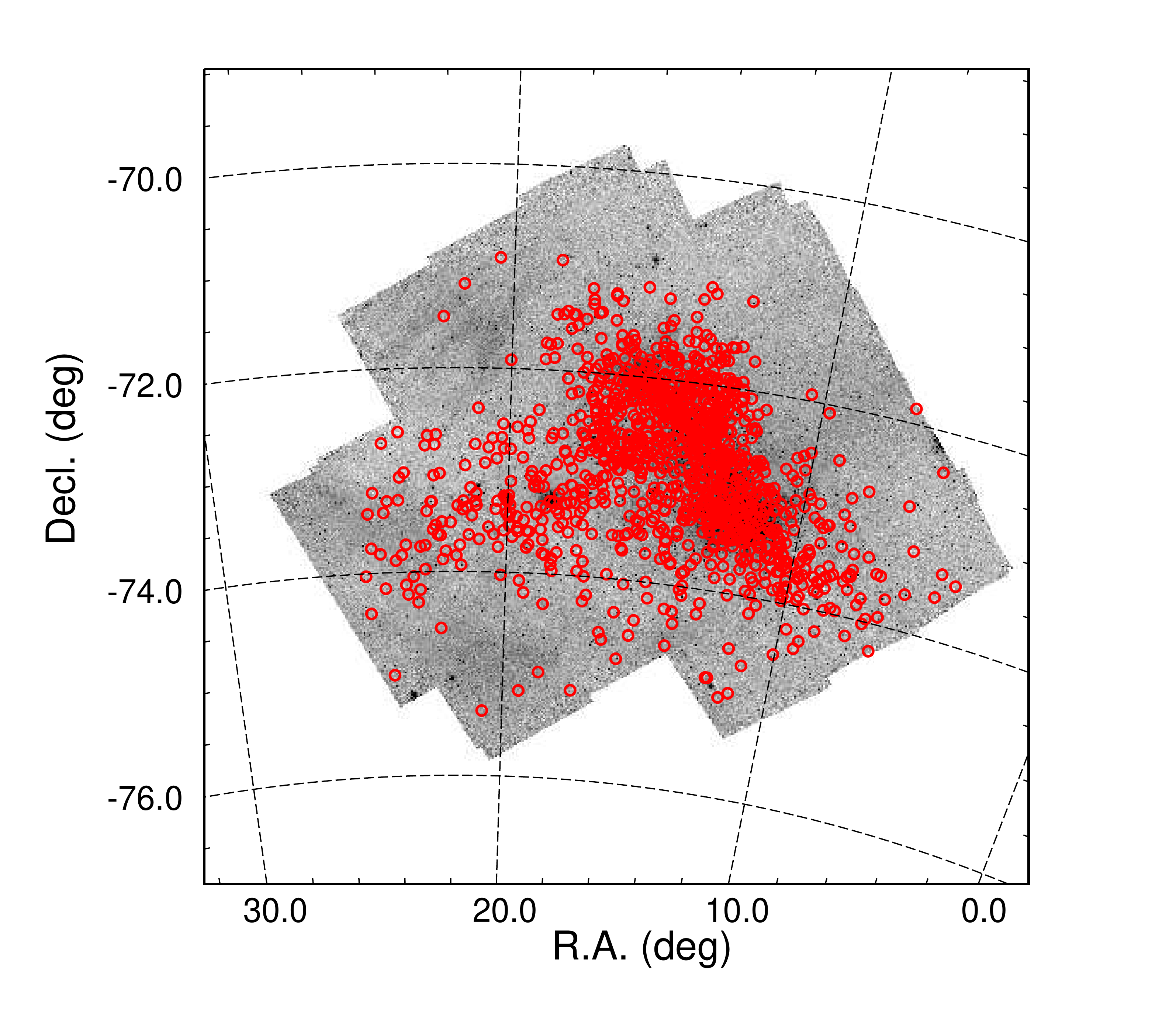}
\caption{Spatial distribution of 1,242 RSG candidates overlapped on the cropped \textit{Spitzer} 8.0 $\mu$m mosaic image (north up and east to the left). Part of the targets are stretching towards the MB (in the direction of south-east) due to the interaction between the LMC and the SMC. \label{spatial}}
\end{figure}

In total, we have 1,242 RSG candidates in our valid RSG sample, with 1,107 targets ranked from 0 to 3, 129 target ranked -1, and 6 targets ranked 4. There are 327 unique spectroscopic RSGs from Simbad and \textit{Spitzer}/IRS after removing the duplications. The same format of datasets as the SMC source catalog were retrieved for the newly added RSG candidates (see Table 3 of \citealt{Yang2019a} regarding to the form and content of the table). Figure~\ref{cmd_gaia_tmass} shows the CMDs of \textit{Gaia} and 2MASS datasets, where the valid RSG sample is overplotted on the SMC source catalog. We use a canonical value of 18.95$\pm0.07$ as the distance modulus of the SMC \citep{Graczyk2014, Scowcroft2016}. However, it is worthwhile to mention that SMC also has a 3D structure, with a full depth of $\sim$0.4 mag \citep{Jacyszyn-Dobrzeniecka2016, Muraveva2018}, which will be discussed later. It appears that the valid RSG sample occupies the most luminous red regions in both CMDs. The extended branch towards the red end (presumably the dusty RSGs) in the \textit{Gaia} CMD is distinctly brighter than the AGBs population. The sample extends to $\rm M_G\approx-4.0~mag$ and $\rm M_{K_S}\approx-6.0~mag$, respectively. Figure~\ref{spatial} shows the spatial distribution of the valid RSG sample overlapped on the cropped \textit{Spitzer} 8.0 $\mu$m mosaic image, for which the stretching of the targets towards the Magellanic Bridge (MB) due to the interaction of the LMC and SMC is obviously seen. We further purified the sample by studying the infrared CMDs and the variability of the objects as described in the next section.

Moreover, we would like to emphasize that, even though our sample size is almost one order of magnitude larger than a decade ago, still it is not fully complete due to several limitations mentioned above. A quick comparison between our sample and a new spectroscopic RSG sample in the SMC (Philip Massey, private communication) indicates that, our main RSG sample (1,405 targets with Rank 0 to 5) and valid RSG sample (1,242 targets with Rank -1 to 4) may reach to about 80\% and 90\% completeness down to $\rm K_S\leq 11.0~mag$, respectively. In that sense, based on the $\sim$80\% completeness of the main sample (there is almost no spectroscopic RSG fainter than $\rm K_S=11.0~mag$), we will expect a total of $\sim$1,800 or more RSGs in the SMC.

\section{Data analysis}
\subsection{2MASS color-magnitude diagrams}

A robust sample is the basis for our study of the RSGs population in the SMC. Since the emission peak of RSGs is close to the near-infrared (NIR) bands due to the low $T_{\rm eff}$, the IR data have smaller variability and are less affected by extinction than the optical data \citep{Chiosi1986, vanLoon1999, Mauron2011}, the NIR and MIR CMDs are used as the first approach to purify and evaluate the RSGs sample. First of all, we considered the 2MASS bands as the primary standard. It is not only because of the reasons mentioned above, but also due to the fact that 2MASS bands are still mainly dominated by the stellar radiation instead of the dust emission, and both the MIST models and the theoretical $\rm J-K_S$ color cuts are available in 2MASS bands. 

\begin{figure*}
\center
\includegraphics[bb=55 405 560 650]{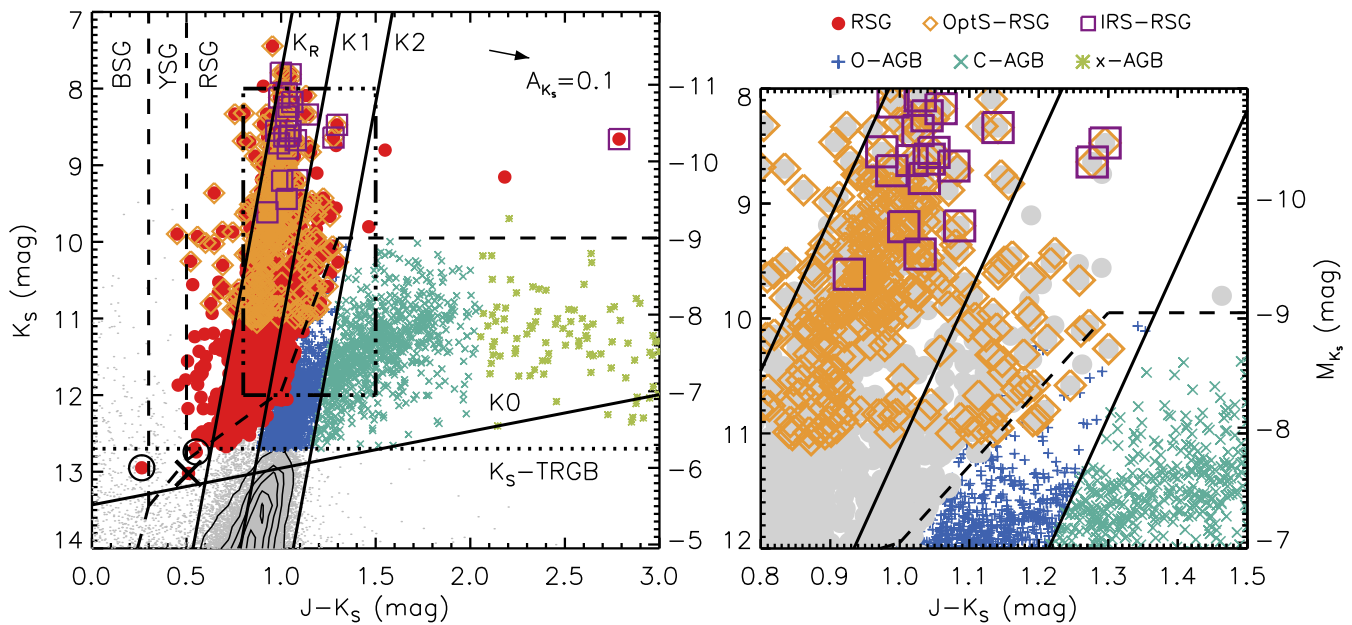}
\caption{$\rm K_S$ versus $\rm J-K_S$ CMDs showing RSG and AGB candidates. Left panel: targets from the SMC source catalog are shown as gray dots with contours indicating the number density (same below), while the 1,242 RSGs candidates are shown as solid circles. The O-AGB (pluses), C-AGB (crosses), x-AGB (asterisks), and RSG populations defined by theoretical $\rm J-K_S$ color cuts are separated by K0, K1, K2, and $\rm K_{R}$ lines (solid lines), respectively (same below). The regions of BSG, YSG, and RSG populations defined by the MIST models are separated by the dashed lines. The optical and MIR spectroscopic RSGs are shown as open diamonds and open squares, respectively, and only represent about 25\% of the whole RSG population. A reddening vector of $\rm A_{K_S}=0.1~mag$ is shown as a reference. Two targets both fainter than the $\rm K_S$-TRGB ($\rm K_S\approx12.7~mag$; dotted line) and IRAC1-TRGB ($\rm IRAC1\approx12.6~mag$) are excluded from the RSG sample, and are marked as big open circles. One heavily obscured target is shown as big cross (same below). The zoom out region is indicated by dashed-doted lines. Right panel: the zoom out region of $\rm 8.0\le K_S \leq 12.0~mag$ and $\rm 0.8 \leq J-K_S \leq 1.5~mag$, where the distribution of the optical spectroscopic RSGs at the red end is following almost exactly the MIST tracks. See text for details.
\label{cmd_tmass}}
\end{figure*}

The left panel of Figure~\ref{cmd_tmass} shows the $\rm K_S$ versus $\rm J-K_S$ diagram of the SMC source catalog (gray dots) overlapped with 1,242 RSG candidates (red solid circles). The foreground extinction of the SMC was not taken into account since it was very small in the 2MASS bands ($\rm A_J\approx$0.03 and $\rm A_{K_S}\approx$0.01 mag), and comparable to the observational uncertainty, if $\rm E(B-V)\sim0.06~mag$ with the Galactic average value of $\rm R_V=3.1$ was adopted \citep{Bessell1991, Gao2013}. Considering $\rm A_{K_S}/A_V\sim0.1$ and the universal IR extinction law \citep{Rieke1985}, a reddening vector of $\rm A_{K_S}=0.1~mag$ is shown for reference by using a precise interstellar dust extinction law from \citet{Wang2019} (it may not show in some diagrams below since it is too small to see). We emphasize that these extinction coefficients are only a good approximation due to the large difference in the metallicity between the SMC and the Milky Way. The regions of different evolved massive star populations (BSGs, YSGs, and RSGs) defined by the MIST models are shown as dashed lines (notice that, the boundaries between RSGs, YSGs, and BSGs are simply defined at $T_{\rm eff}=5,000$ and 7,500 K, respectively; see also Figure 13 and 14 of \citealt{Yang2019a}). The regions of Oxygen-rich AGBs (O-AGBs; K$_{\rm S}$ $<$ K$_{\rm S}$-TRGB ($\approx$12.7 mag, dotted line; \citealt{Cioni2000}) and K1 $<\rm J-K_{S} <$ K2), Carbon-rich AGBs (C-AGBs; K$_{\rm S}$ $<$ K$_{\rm S}$-TRGB and K2 $<\rm J-K_{S} <$ 2.1 mag), extreme-AGBs (x-AGBs; $\rm J-K_{S}>2.1$ mag), and RSGs ($\rm K_{R}$; $\rm \Delta(J-K_{S})=0.25$ mag from the O-AGBs) defined by the theoretical $\rm J-K_S$ color cuts are shown as solid lines (all AGBs are brighter than the K0 line except x-AGBs; the distance and 0.05 mag for the metallicity between the LMC and the SMC are corrected; \citealt{Cioni2006b}; see more details in Section 3 of \citealt{Boyer2011} and Section 4 of \citealt{Yang2019a}).

At the bright end, it can be seen that all the spectroscopic RSGs are brighter than $\rm K_S\approx11.0~mag$ and only represent about 25\% of the whole population. It is not even fully representative for the bright population since there are around six hundreds target with $\rm K_S\lesssim11.0~mag$. This prevents us from constraining the RSGs population solely based on the incomplete sample of spectroscopic RSGs, until the next generation of large-scale spectroscopic data are obtained (e.g., 4MOST, MOONS; \citealt{Cirasuolo2012, deJong2012}). We also would like to emphasize that, due to the strict constraints applied to our source catalog and also the limitations of different catalogs (e.g., saturation limits, photometric quality cuts, sky coverage and so on), there may still be a very small chance that we miss a few targets at the very bright end (e.g., $\rm K_S\sim8.0$ mag; most likely due to the saturation problem of \textit{Spitzer}). However, since a comparison between our valid sample and a quick crossmatching of the 2MASS point source catalog (which is less constrained and also shallower than our source catalog) and \textit{Gaia} DR2 with relaxed criteria indicated that the bright ends of two samples were identical, we concluded that we did not lose any of the brightest RSGs.

At the faint end, we excluded targets fainter than both the $\rm K_S$-TRGB ($\rm K_S\approx12.7~mag$) and IRAC1-TRGB ($\rm IRAC1\approx12.6~mag$; \citealt{Boyer2011}), to eliminate possible contamination from the red giant branch stars (RGBs), since our sample reached down to the TRGB. Two targets were identified, which resulted in a reduced sample of 1,240 RSG candidates. By checking the TRGBs in both $\rm K_S$ and IRAC1 bands, we assured that the obscured targets were not excluded \citep{Boyer2011}. However, notice that, as we do not include the majority of RSG candidates with Rank 4 or 5, there is still a chance that some of them may be true RSGs. Furthermore, even if we constrain our sample to be brighter than the TRGB, as shown in Figure 15 of \citet{Yang2019a}, there are still a bunch of targets from the RSG branch that extend towards fainter magnitude. Especially, there are hundreds of targets clumping at the region between RSGs, RGBs, AGBs, and red clump stars (RCs). Currently, the nature of these fainter targets is unknown, but could be a mixture of different populations including very faint RSGs.

For the blue end of $\rm J-K_S$ color, we expect there to be overlapping between the RSGs and YSGs, which is due to three reasons: First, there is no clear boundary between RSGs and YSGs. Second, some RSGs can change spectral type from early-K to late-G due to variability. Third, the average spectral type of RSGs shifts towards earlier types at lower metallicities \citep{Levesque2006, Massey2007, Gonzalez2015, Dorda2016}. 

For the red end of $\rm J-K_S$ color, as described in the Section 4 of \citet{Yang2019a}, there was a discrepancy between the MIST models and the theoretical $\rm J-K_S$ color cuts, where a significant number of RSGs candidates (206 targets) selected by the MIST models were located within the O-AGBs region defined by the theoretical color cuts. Among them, $\sim$23\% (47 targets) are spectroscopic RSGs. This poses a real dilemma of defining the RSG population. On the one hand, considering that the majority of the spectroscopic RSGs are brighter than $\rm K_S\approx11.0~mag$, the ratio of the spectroscopic RSGs is as high as $\sim$43\% (46/106; with $\rm K_S\leq11.0~mag$) for the RSG candidates in the O-AGBs region. Moreover, the right panel of Figure~\ref{cmd_tmass} shows the zoom out region of $\rm 8.0\leq K_S\leq 12.0~mag$ and $\rm 0.8\leq J-K_S\leq 1.5~mag$, where the distribution of the optical spectroscopic RSGs at the red end is following almost exactly the MIST tracks, indicating that the MIST models may be correct to a certain extent, since there is no spectroscopic RSG in the fainter magnitude up to now. On the other hand, many spectroscopic classifications are based on line ratios or morphological classifications \citep{Dorda2018}, that are a continuum from AGBs to RSGs, and both the RSGs and AGBs show spectral variability \citep{Bessell1996, Dorda2016}. This probably means that spectroscopic identification may not be a golden standard, as the difference between AGBs and RSGs is not always evident when looking at a spectrum, especially at the overlapping region of the CMD.

\begin{figure*}
\center
\includegraphics[bb=55 405 560 655]{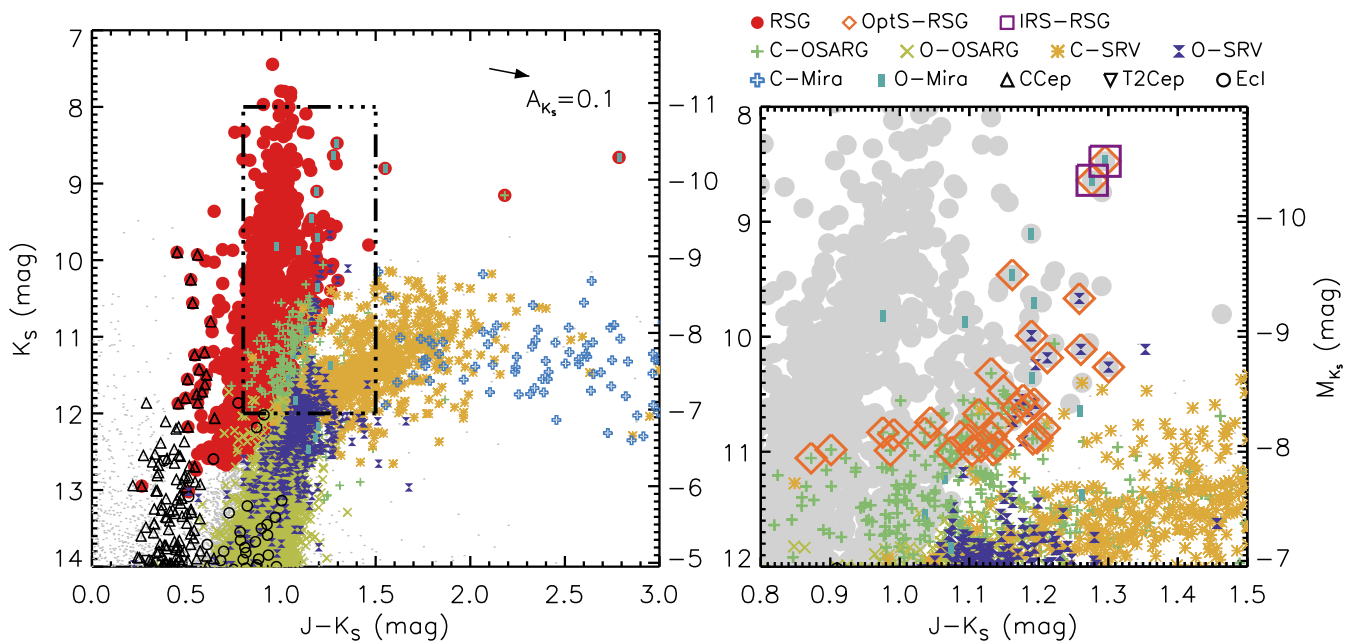}
\caption{Right panel: the same diagram as Figure~\ref{cmd_tmass} but with variable classifications from OGLE. Different types of variables are shown as different symbols. An overlapping of OSARGs, SRVs, and Miras can be seen at the lower half of the zoom out region. Left panel: the same zoom out region as Figure~\ref{cmd_tmass}, where 24 C-OSARGs, 7 O-SRVs and 3 O-Miras are optical spectroscopic RSGs (open diamonds). More interestingly, two O-Miras at the upper right are classified as both optical and MIR spectroscopic RSGs (open squares).
\label{cmd_tmass_ogle}}
\end{figure*}

For further analysis, the left panel of Figure~\ref{cmd_tmass_ogle} shows the same diagram of $\rm K_S$ versus $\rm J-K_S$ with variable classifications \citep{Soszyski2011, Soszyski2015, Pawlak2016, Soszyski2018} from the Optical Gravitational Lensing Experiment (OGLE; \citealt{Udalski1992, Szymanski2005, Udalski2008, Udalski2015}), and are matched with our source catalog and RSG sample within $1''$ (the numbers and classifications are listed in Table~\ref{var_ogle}). Initially, it appears that many Carbon-rich OGLE Small Amplitude Red Giants (C-OSARGs) are located within the O-AGBs region between K1 line and the red boundary of MIST models and overlapped with our RSG sample. Meanwhile, the same zoom out region (the right panel of Figure~\ref{cmd_tmass_ogle}) indicates that 24 C-OSARGs, 7 Oxygen-rich Semiregular Variables (O-SRVs) and 3 Oxygen-rich Miras (O-Miras) are optical spectroscopic RSGs. When considering targets with $\rm K_S\leq11.0~mag$, the ratios of spectroscopic RSGs for C-OSARGs, O-SRVs, and O-Miras are about one-third ($\sim$34\%; 22/64), half ($\sim$54\%; 7/13), and one-fourth (25\%; 3/12), respectively. Still, similar to the spectral conundrum, the variable classification also has the same problem. It was mainly done based on the light curve (LC) morphology, position of a star in the period-luminosity, color-magnitude, and period-amplitude diagrams \citep{Soszyski2011}. Such process was a complex issue, due to the complicated nature of the LCs, which often showed multiperiodicity, irregular variations, changes of the mean magnitudes, or modulations of periods, phases, and amplitudes \citep{Soszyski2009}. In that sense, there is again a continuum with similarity and overlapping from OSARGs to SRVs to Miras, which also can be seen from the both panels of Figure~\ref{cmd_tmass_ogle}. It is possible that some medium- or low-luminosity RSGs, which are supposedly to be SRVs, are misclassified as OSARGs or Miras, but, vice versa. The LCs of some confused targets were visually inspected. However, based on our previous experiences of RSGs LCs (e.g., \citealt{Kiss2006, Yang2011, Yang2012, Soraisam2018, Ren2019}), it is hard to say whether these targets are truly RSGs or not. More interestingly, two O-Miras at the upper right of the zoom out region are classified as both optical and MIR spectroscopic RSGs.

In brief, there is a blurred boundary between AGBs and RSGs, and no efficient way to distinguish them by using current data. More data (e.g., spectra with higher resolution and larger wavelength coverage, narrow-filter photometries, multiwavelength time-series observations, sub-millimeter data) are strongly needed to further distinguish between AGBs and RSGs.

Meanwhile, based on the fact that obscured objects would be brighter and redder at the longer wavelengths compared to the shorter wavelengths, we identified one heavily obscured target (big cross) in the entire RSG sample (from Rank -1 to 5) by adopting $\rm K_S>12.7~mag$ or $\rm IRAC1>12.6~mag$ (targets were fainter either than the $\rm K_S$-TRGB or IRAC1-TRGB), $\rm J-IRAC4\geq3.0~mag$, and $\rm IRAC4<10.0~mag$ (see also next subsection). The obscured target shows high luminosities at the longer wavelengths dominated by the dust emission but dimming at the shorter wavelengths, in particular it is located below the $\rm K_S$-TRGB. Although it lies in the x-AGBs region at the longer wavelengths, the obscured target is found very close to the RSGs region in the 2MASS CMD and inside the RSGs regions in other CMDs (e.g., Gaia, SkyMapper, NSC, which are not shown here), for which it will move along the opposite direction of the reddening vector without the extinction (to be brighter and bluer). Other RSG candidates located in the AGBs regions at the longer wavelengths are similar to the obscured target, that they will be located in the RSG region without the extinction.

\subsection{Mid-infrared color-magnitude diagrams}

\begin{figure*}
\center
\includegraphics[bb=125 365 505 690, scale=0.65]{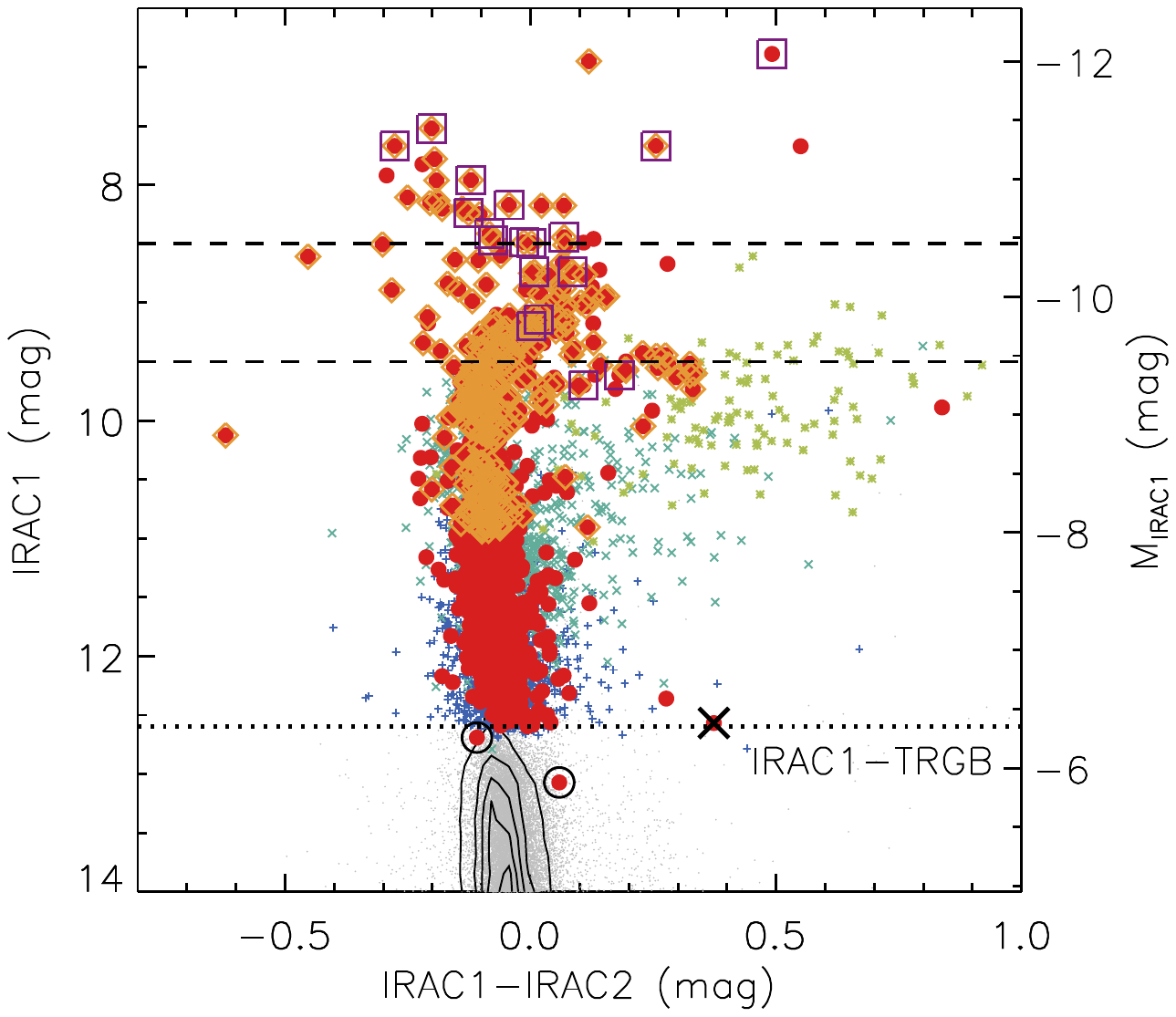}
\includegraphics[bb=125 365 505 690, scale=0.65]{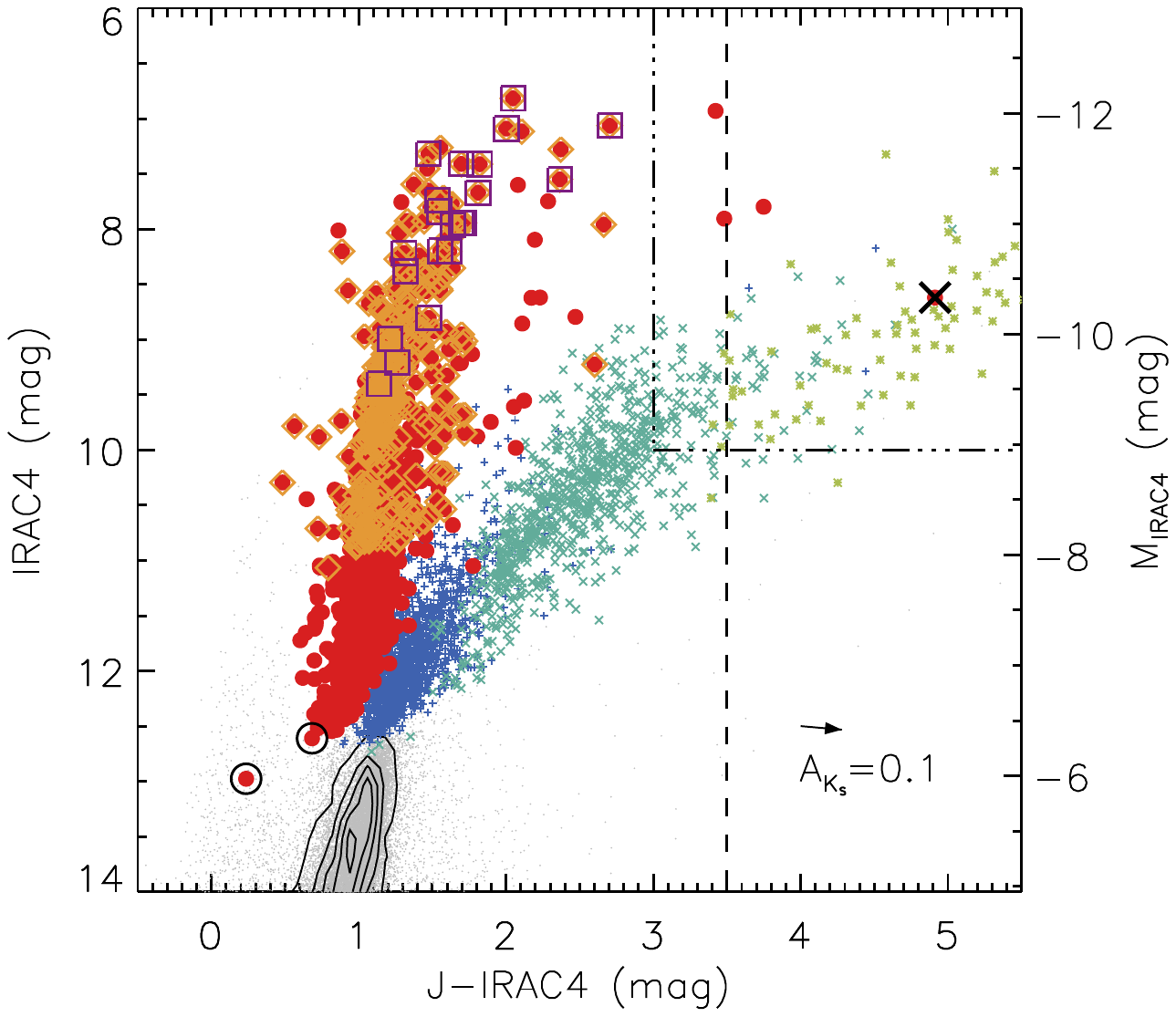}
\includegraphics[bb=125 365 505 690, scale=0.65]{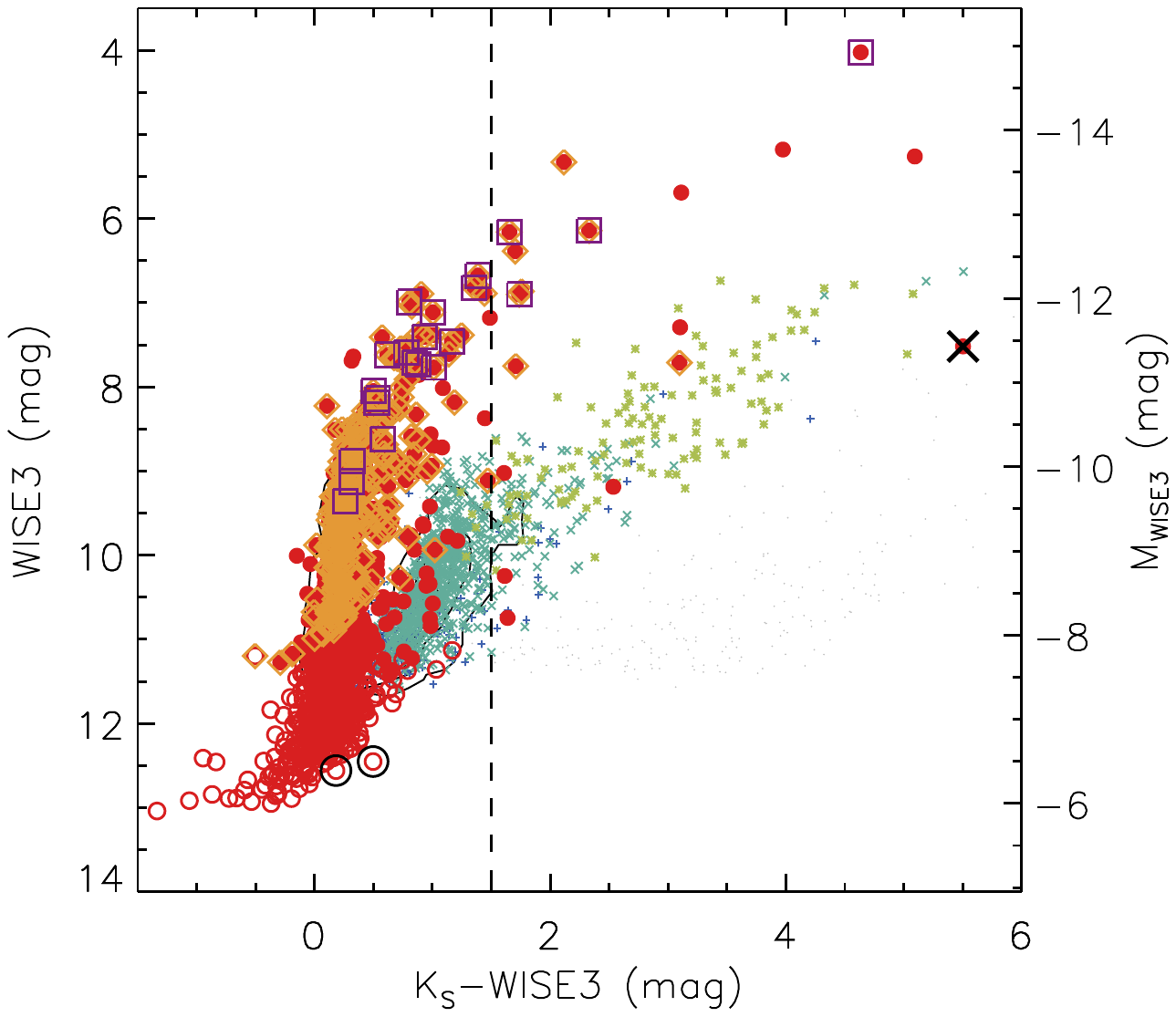}
\includegraphics[bb=125 365 505 690, scale=0.65]{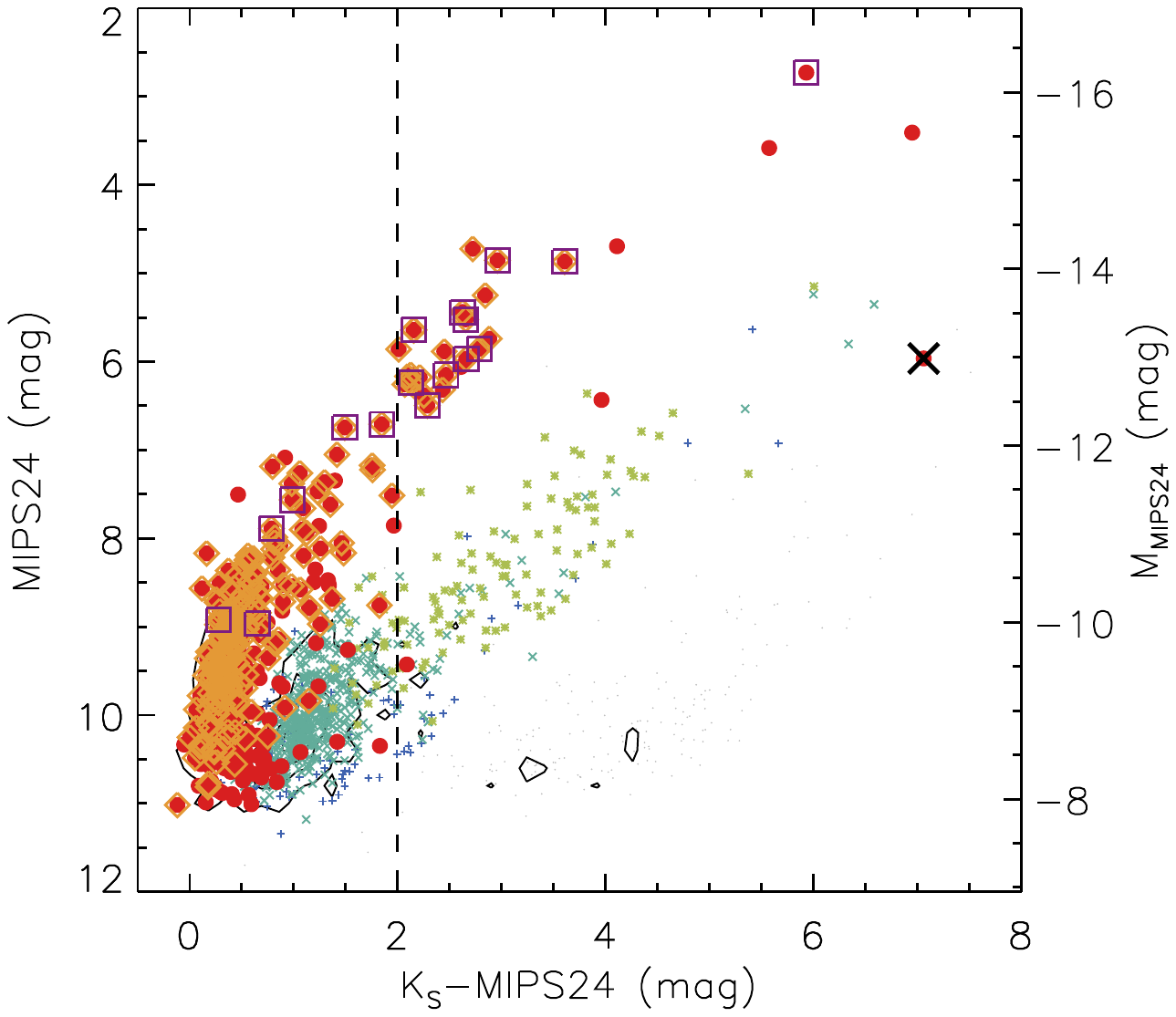}
\caption{Upper left: IRAC1 versus $\rm IRAC1-IRAC2$ diagram. The RSG sample mainly shows negative value of $\rm IRAC1-IRAC2$ due to the CO absorption around 4.6 $\mu$m. In particular, it turns bluewards ($\rm IRAC1 \gtrsim 9.5~mag$) with CO absorption, redwards ($\rm 8.5 \lesssim IRAC1 \lesssim 9.5$) with enhanced mass loss due to both luminosity and variability, and again bluewards ($\rm IRAC1 \lesssim 8.5$) with PAH emission, along with the increasing of luminosity. There is another fainter branch ($\rm 9.3\lesssim IRAC1 \lesssim10.0~mag$ and $\rm 0\lesssim IRAC1-IRAC2 \lesssim 0.35~mag$) extended to the x-AGBs region with unknown origin. Upper right: IRAC4 versus $\rm J-IRAC4$ diagram. The dashed-dotted lines region indicates the selection criteria for the obscured target. Bottom left: WISE3 versus $\rm K_S-WISE3$ diagram. For clarity, RSG candidates with $\rm S/N_{WISE3}<10$ are shown as red open circles, while background targets from SMC source catalog with $\rm S/N_{WISE3}<10$ are not shown in the diagram. The RSGs population shows redder color of $\rm J-IRAC4$ and $\rm K_S-WISE3$ along with the increasing of IRAC4/WISE3 luminosity. Bottom right: MIPS24 versus $\rm K_S-MIPS24$ diagram. The majority of the MIPS24 detected targets are also spectroscopic RSGs. The brightest RSGs candidates form a parallel sequence with respect to the x-AGBs population. The vertical dashed lines indicate the approximate blue boundaries of x-AGBs. See text for details. 
\label{cmd_jkiw}}
\end{figure*}

Figure~\ref{cmd_jkiw} shows four CMDs, with various combinations of mid-IR and near-IR filters. For IRAC1 versus $\rm IRAC1-IRAC2$ diagram, the RSG sample mainly shows negative color index of $\rm IRAC1-IRAC2$ with median value of -0.076, due to the CO absorption around 4.6 $\mu$m \citep{Verhoelst2009, Britavskiy2015, Reiter2015}. In particular, the main population of RSG sample ($\rm IRAC1 \gtrsim 9.5~mag$ and $\rm IRAC1-IRAC2 \lesssim 0~mag$; $\sim$80\%) shifts slightly bluewards as the luminosity increases, indicating possibly the strengthening of CO absorption. However, this trend turns redwards when targets brighter than $\rm IRAC1\approx9.5~mag$ (dashed line), which may be due to the enhanced mass loss caused by both luminosity and variability \citep{Yang2018}. At the bright end ($\rm IRAC1\lesssim8.5~mag$; dashed line), the trend turns again bluewards, which may be explained by the the polycyclic aromatic hydrocarbon (PAH) emission in 3.3 $\mu$m captured by the IRAC1 filter \citep{Buchanan2006, Verhoelst2009}. More importantly, it seems that only the very bright RSGs will produce PAH, although it may need further confirmation from the MIR spectroscopy. Notably, there is another branch of targets (also containing many spectroscopic RSGs) at fainter magnitude ($\rm 9.3\lesssim IRAC1 \lesssim10.0~mag$ and $\rm 0\lesssim IRAC1-IRAC2 \lesssim 0.35~mag$) extended to the x-AGBs region. This branch may be due to the complicated interaction between PAH, CO, dust, and/or unknown chemical composition, or, as discussed in previous section, due to the misclassification of AGBs.

In the IRAC4 versus $\rm J-IRAC4$ diagram, the $\rm J-IRAC4$ color index is mainly affected by the infrared excess related to the broad 9.7 $\mu$m silicate emission feature, which is attributed to the Si-O stretch resonance, and/or the relatively weak 7.6 $\mu$m~PAH emission feature captured by the IRAC4 band \citep{Sloan2008, Bonanos2009, Boyer2011}. Meanwhile, for WISE3 versus $\rm K_S-WISE3$ diagram (for clarity, RSG candidates with WISE3-band signal to noise ratio (S/N) less than 10 are shown as red open circles, while background targets from SMC source catalog with $\rm S/N_{WISE3}<10$ are not shown in the diagram), the $\rm K_S-WISE3$ color index is also affected by the 9.7 $\mu$m feature, and/or 18 $\mu$m silicate emission from Si-O-Si bending mode in the SiO$_4$ tetrahedron \citep{Josselin2000, Verhoelst2009, Chen2016}. Both of them can be considered as proxies for the MLR. Since oxygen-rich dust shows silicate features at 9.7 and 18 $\mu$m that appear in emission at low optical depths and absorption at high optical depths (i.e., with high MLR; \citealt{Sylvester1999, Kemper2000}), it also explains why the obscured and some other RSG candidates are located in the AGBs region at the longer wavelengths (targets with high MLR will become red and faint). In any case, the RSG population shows redder colors of $\rm J-IRAC4$ and $\rm K_S-WISE3$ along with the increasing of IRAC4 and WISE3 luminosities, indicating the growth of MLR and circumstellar envelope. 

In the MIPS24 versus $\rm K_S-MIPS24$ diagrams, since the MIPS24 band traces the relatively cold dust continuum with very little contribution from the stellar photosphere, $\rm K_S-MIPS24$ color is also a good indicator of MLR. Due to the larger distance compared to the LMC and lower sensitivity at the longer wavelength, the majority of the RSGs candidates remain undetected in MIPS24 band ($\sim$67\%) or only detected with an upper limit in WISE4 band ($\sim$76\%), while most of the detected targets are also spectroscopic/bright RSGs. From the diagram, it can been seen that there are two parallel sequences showing ascending trends in both MIPS24 luminosity and $\rm K_S-MIPS24$ color, formed by the brightest RSG candidates and x-AGBs, respectively. This likely indicates a similar large MLR in these stars.

Furthermore, by simultaneously inspecting the last three diagrams, it occurs to us that the brightest RSG candidates are likely having high MLR comparable to the x-AGBs, as x-AGB stars may be experiencing a ``superwind'' with extreme MLR and a thick dust envelope \citep{vanLoon2006, Boyer2011, Hofner2018}. This can be seen by comparing the colors (infrared excesses) of the brightest RSG candidates and x-AGB population. For example, in the IRAC4 versus $\rm J-IRAC4$ diagram, the x-AGBs are redder than $\rm J-IRAC4\approx3.5~mag$, where only two targets from our sample have the similar colors. Proceeding to longer wavelengths, the number of brightest RSG candidates similar to x-AGBs' color increases to 18 ($\rm K_S-WISE3\gtrsim1.5~mag$) and 30 ($\rm K_S-MIPS24\gtrsim2.0~mag$), respectively. The inconsistency between infrared excesses of the relatively shorter (8 $\mu$m) and longer (12 and 22 $\mu$m) wavelengths suggests that, it may not be enough to characterize the MLR of brightest RSGs even with the wavelength as long as 8 $\mu$m.

In total, there were 1,240 targets remaining in our RSG sample. No further constraints were made on the CMDs as we did in \citet{Yang2018}. This is due to three reasons. First, all the RSG candidates have already been constrained either by the astrometric parameters and the evolutionary models, or the radial velocities and spectral/luminosity types, which indicates that their luminosities and colors are well determined and there is almost no foreground contamination. Second, there are too many combinations of filters (50 optical and infrared filters) and not all the combinations are efficient, as the completeness and sky coverage are different for different datasets. Third, the optical filters suffer more from extinction and reddening than the infrared filters, which is even worse for the targets with larger MLR.

\subsection{Color-Color Diagrams}

We also investigated the distribution of RSGs population on the color-color diagrams (CCDs). Generally speaking, stars will move along the direction of reddening vector due to the interstellar reddening but deviate from it due to the circumstellar reddening \citep{Bonanos2010, Wachter2010, Messineo2012}. 

\begin{figure*}
\center
\includegraphics[bb=125 365 505 690, scale=0.65]{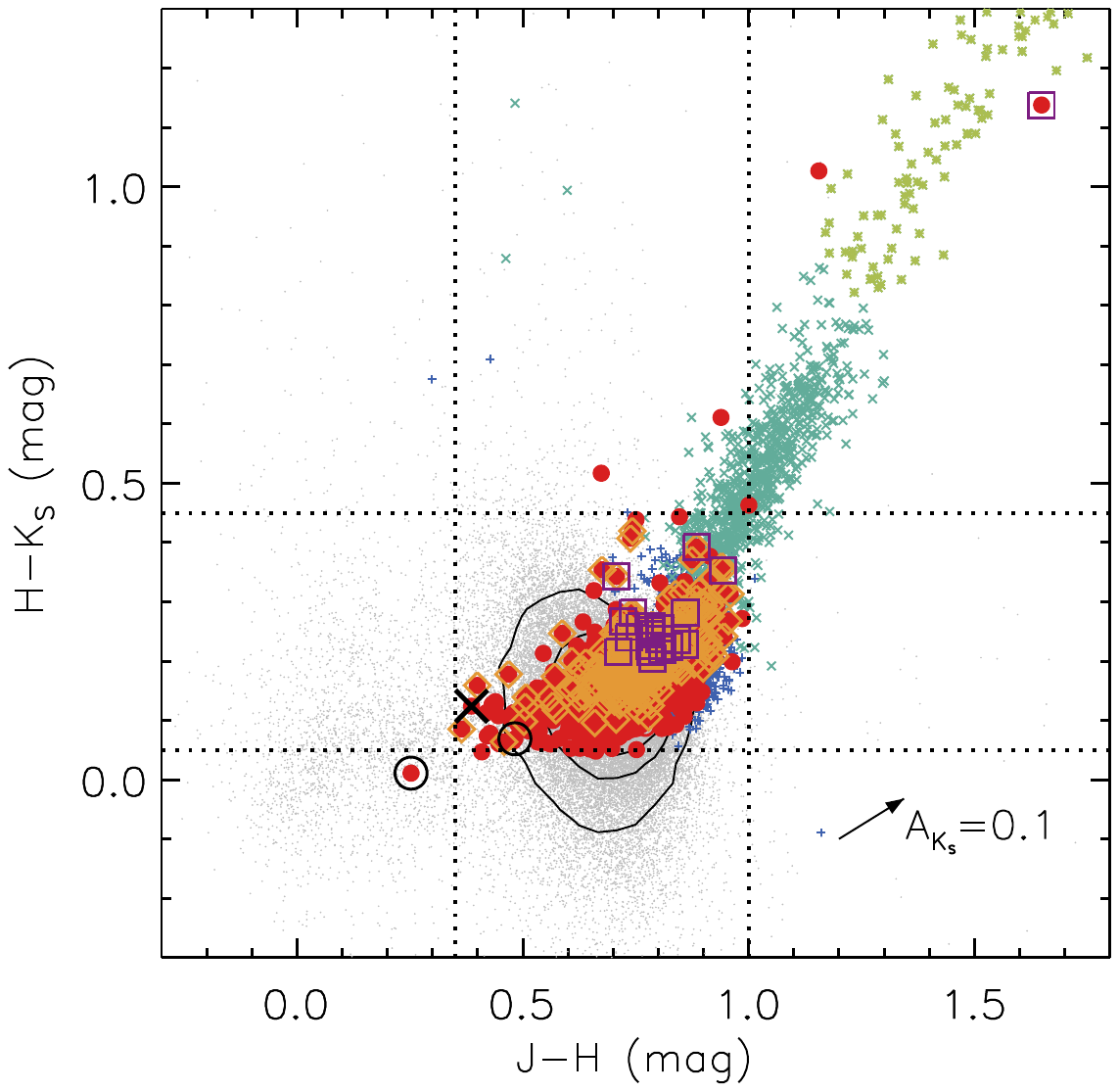}
\includegraphics[bb=125 365 505 690, scale=0.65]{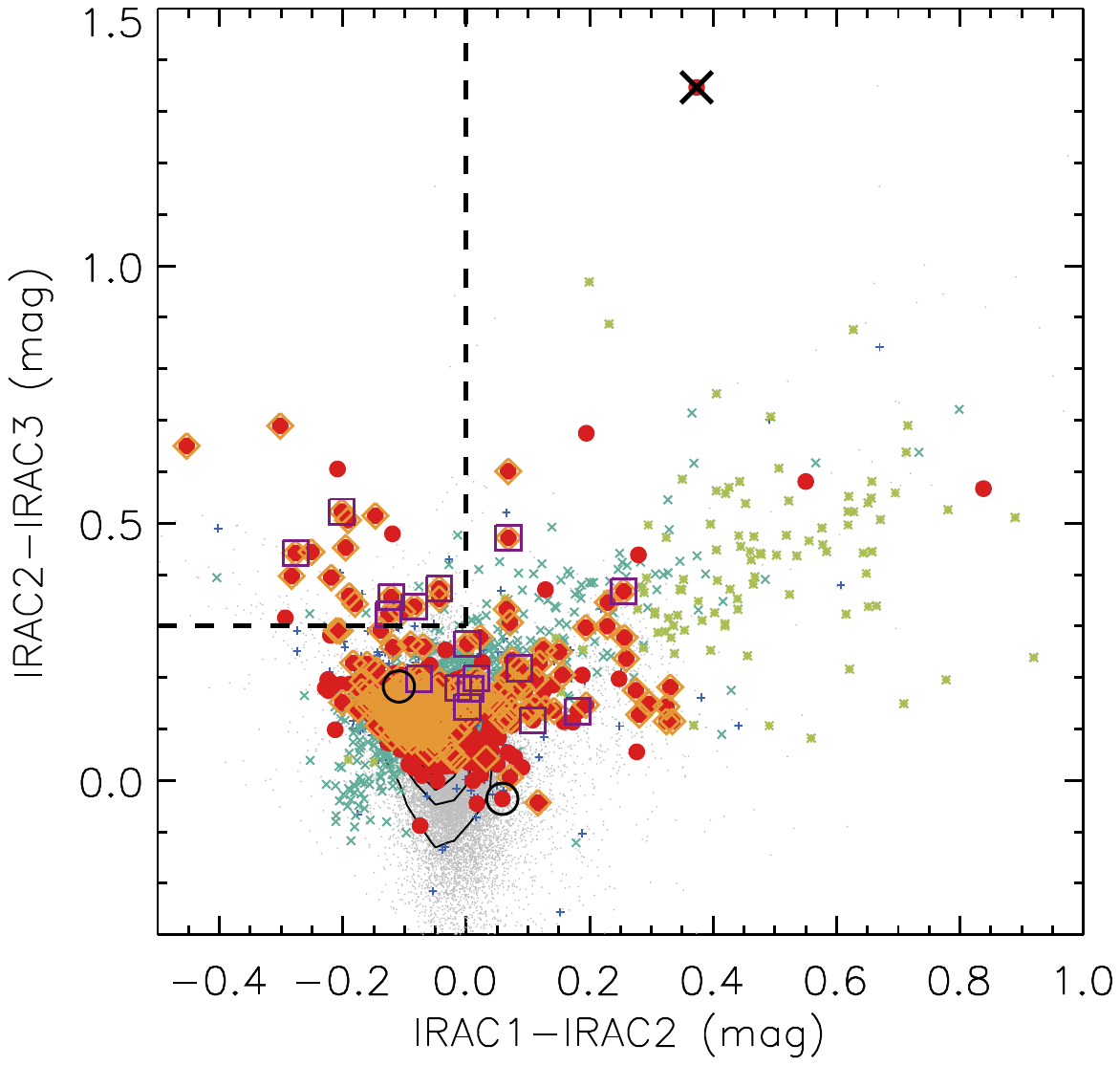}
\includegraphics[bb=125 365 505 690, scale=0.65]{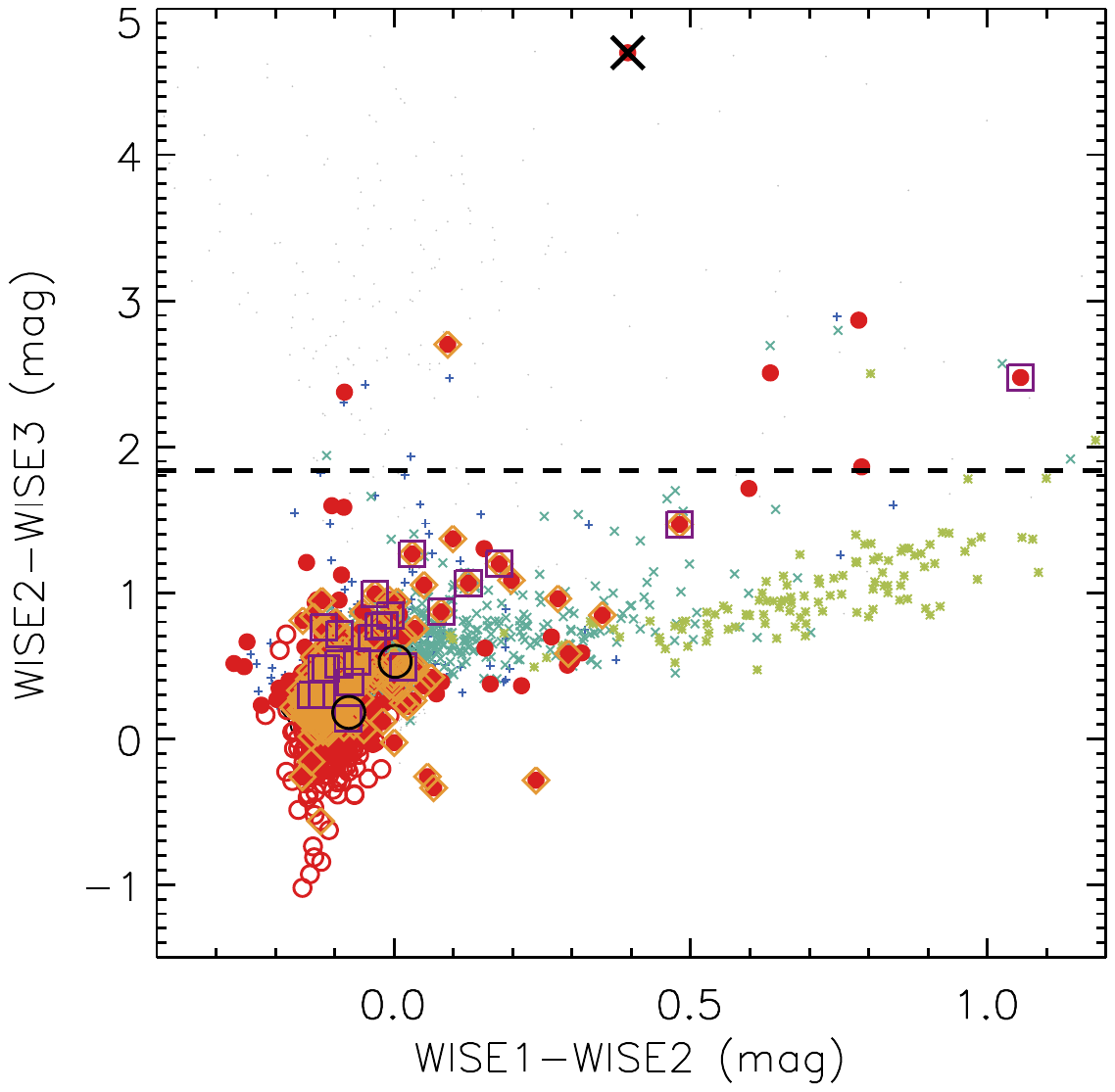}
\includegraphics[bb=125 365 505 690, scale=0.65]{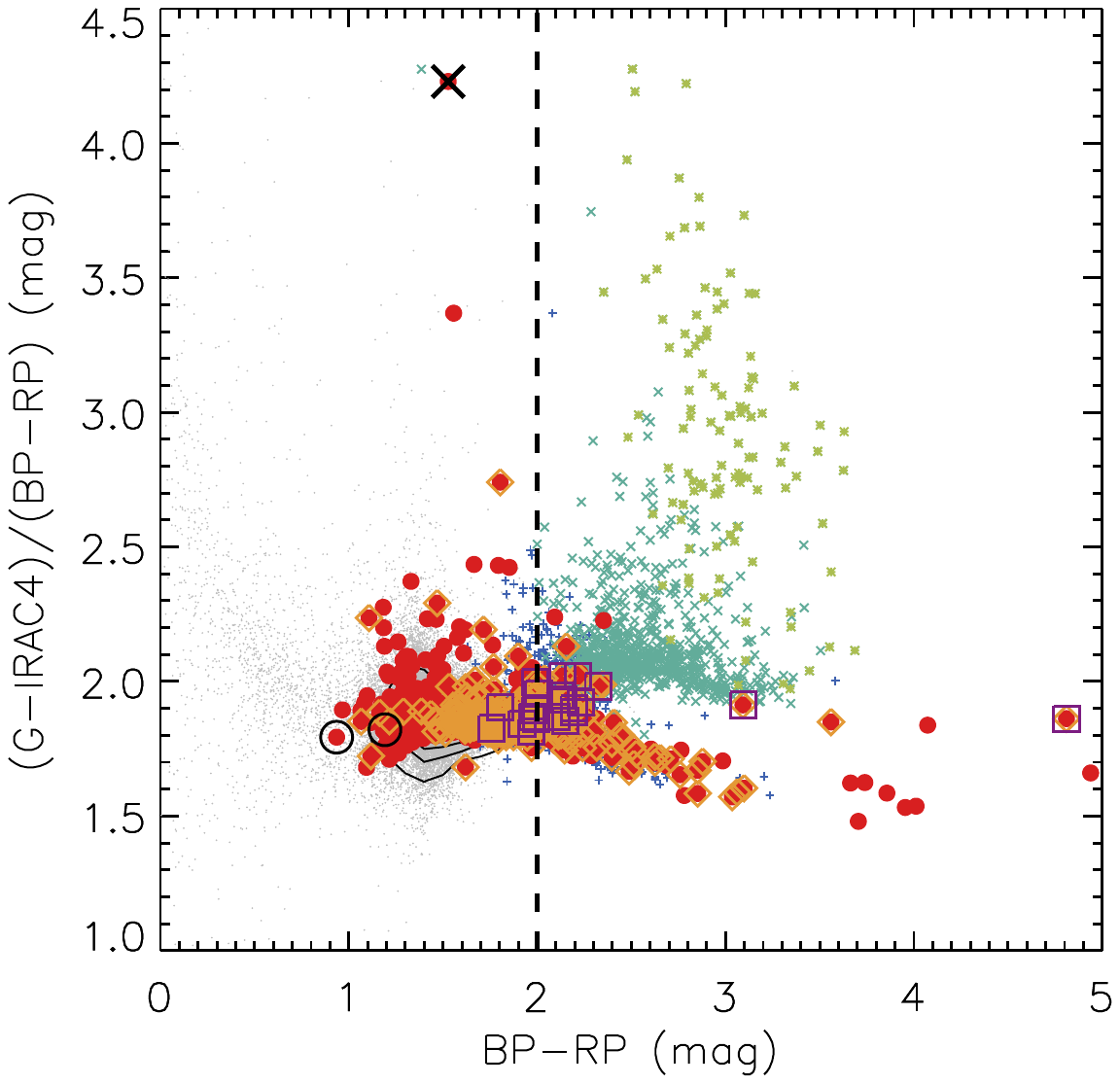}
\caption{Upper left: 2MASS CCD of $\rm J-H$ versus $\rm H-K_S$. The RSGs population is clumping within the range of $\rm 0.35 \lesssim J-H \lesssim 1.0$ and $\rm 0.05 \lesssim H-K_S \lesssim 0.45$ (dotted lines), and is indistinguishable from the O-AGBs. Upper right: IRAC CCD of $\rm IRAC1-IRAC2$ versus $\rm IRAC2-IRAC3$. Previously defined PAH emission region of $\rm IRAC1-IRAC2\leq0$ and $\rm IRAC2-IRAC3\geq0.3$ is shown as dashed lines \citep{Yang2018}. There are only about only about 4\% (after correcting for the completeness) of RSG candidates show PAH emission features. Bottom left: WISE CCD of $\rm WISE1-WISE2$ versus $\rm WISE2-WISE3$. The horizontal dashed line indicates $\rm F_{\nu,WISE2}=F_{\nu,WISE3}$ ($\rm WISE2-WISE3=1.84$), for which only seven targets above it. Bottom right: optical-MIR color excesses of $\rm BP-RP$ versus $\rm (G-IRAC4)/(BP-RP)$. RSG candidates (and O-AGBs), C-AGBs, and x-AGBs having lowest, moderate, and highest ratio of $\rm (G-IRAC4)/(BP-RP)$ at $\rm BP-RP\geq 2.0$ (vertical dashed line), respectively.
 \label{ccd}}
\end{figure*}

The upper left panel of Figure~\ref{ccd} shows the 2MASS CCD of $\rm J-H$ versus $\rm H-K_S$. Even with our extended sample size, it is still the same as in \citet{Yang2018} and previous studies. The RSGs population clumps within the range of $\rm 0.35 \lesssim J-H \lesssim 1.0$ and $\rm 0.05 \lesssim H-K_S \lesssim 0.45$ (dotted lines), which is in agreement with \citet{Rayner2009}. Moreover, it is indistinguishable from the O-AGBs due to the similar intrinsic temperatures and chemical compositions as well as the low-luminosity RGBs, while C-AGBs and x-AGBs are distinguishable and show different tendencies due to different chemical composition and/or MLRs.

The upper right and bottom left panels of Figure~\ref{ccd} show IRAC ($\rm IRAC1-IRAC2$ versus $\rm IRAC2-IRAC3$) and WISE ($\rm WISE1-WISE2$ versus $\rm WISE2-WISE3$) CCDs, respectively. For both IRAC and WISE CCDs, as expected, the majority of RSG candidates show blue color (<0) in $\rm IRAC1-IRAC2$ and $\rm WISE1-WISE2$ due to CO absorption as mentioned before. In $\rm IRAC1-IRAC2$ versus $\rm IRAC2-IRAC3$ diagram, we indicated previously defined PAH emission region of $\rm IRAC1-IRAC2\leq0$ and $\rm IRAC2-IRAC3\geq0.3$ shown as dashed lines \citep{Yang2018}. However, for the SMC, only about 4\% (after correcting for the completeness, which is based on the estimation from the results of \citealt{Yang2018} and our source catalog of the LMC; \citealt{Yang2019b}) of RSG candidates show PAH emission features, which is much lower than the Milky Way and LMC ($\sim$15\%; \citealt{Verhoelst2009, Yang2018}). This is likely consistent with the strong metallicity dependence of PAH abundance due to the shattering of carbonaceous grains being the source of PAHs, which may be caused by the harder and more intense radiation fields in low-metallicity environments, and/or the enhanced SN activity \citep{Engelbracht2005, OHalloran2006, Draine2007, Seok2014, Shivaei2017}. Still, most of the scenarios are presumably applied on a much larger scale dominated by the interstellar medium (ISM), according to the observation of nearby and high-redshift galaxies. Meanwhile, the mechanism for individual star, especially for PAH in O-rich RSGs, is still unclear. Moreover, since the presence of PAHs in the ISM could not be explained entirely by the C-AGBs \citep{Matsuura2013}, the contribution of RSGs might be considered, and further MIR spectroscopy is needed to understand the scenario. For $\rm WISE1-WISE2$ versus $\rm WISE2-WISE3$ diagram, compared to the LMC, there are only seven targets showing extreme MIR excess in terms of $\rm WISE2-WISE3\ge1.84$ ($\rm F_{\nu,WISE2}=F_{\nu,WISE3}$; dashed line). It is likely due to the metallicity dependence of dust-driven outflows, as a lower metallicity may result in a smaller mass of dust grains, and consequently, a smaller integrated cross-section for radiation pressure on dust grains and a lower MLR \citep{Justtanont1992, Marshall2004, Mauron2011}. 

Finally, the bottom right panel of Figure~\ref{ccd} shows optical-MIR color excesses of $\rm BP-RP$ versus $\rm (G-IRAC4)/(BP-RP)$, which simply reflects the relation between the photosphere/spectral type ($\rm BP-RP$) and the reddening/dust excess ($\rm G-IRAC4$). It can be seen from the diagram that, targets with $\rm BP-RP\geq 2.0$ are split into three groups, where RSG candidates (and O-AGBs), C-AGBs, and x-AGBs having lowest, moderate, and highest ratio of $\rm (G-IRAC4)/(BP-RP)$, respectively, which may be caused by the different chemical composition and/or large MLR.

\subsection{Mid-infrared variability}

As stated in \citet{Yang2018} and other studies \citep{Wood1983, Kiss2006, Groenewegen2009, Yang2011, Yang2012}, RSGs typically have smaller variability in the same magnitude range compared to AGBs, which can be used as an indicator to separate them. We analyzed the long-term epoch-binned MIR time-series data from WISE for our RSG sample with 1,240 targets. For each target, the median value of single-epoch MADs (approximately five to ten days for each epoch) was also calculated as an estimation of short-term variability. Among them, 48 targets ($\sim$4\%) do not have MIR variability statistics, which is likely due to the quality cuts applied when retrieving the time-series data (see more details about the binning method and WISE photometric quality cuts in Section 3 of \citealt{Yang2018}). Even though our SMC source catalog included other variability statistics, they were neither relatively complete in the spatial or magnitude range, nor homogeneous in the sampling, which left the WISE data as the only choice we had. The saturation problem for WISE data was discussed in the Section 4.4 of \citet{Yang2018}, which was more mitigated in the SMC due to the larger distance compared to the LMC.

\begin{figure*}
\center
\includegraphics[bb=65 415 535 700]{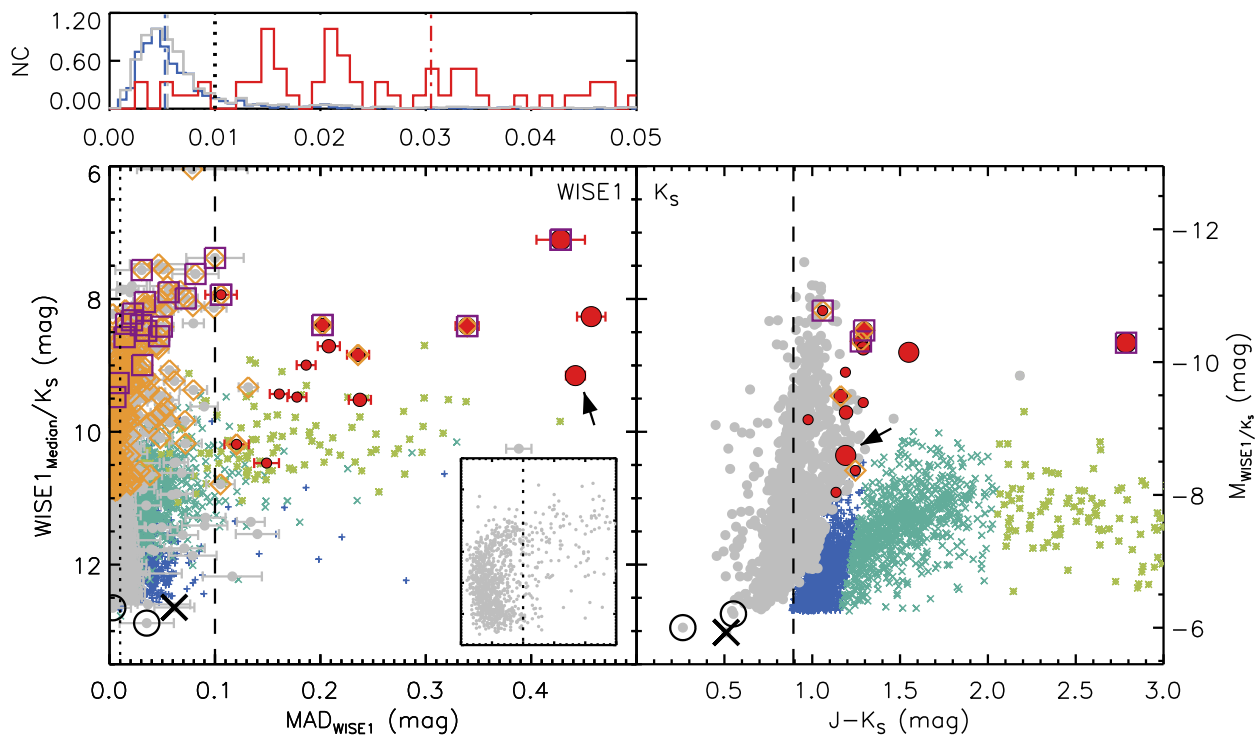}
\caption{Left: MAD versus median magnitude in WISE1 bands. For convenience, the diagram only shows the RSG sample and AGB populations. The median value of single-epoch MADs (each epoch is approximately five to ten days) of each target is shown as the error. The optical and MIR spectroscopic RSGs are shown as open diamonds and open squares, respectively. The MIR variability of RSG sample increases along with luminosity. In total, $\sim$21\% of targets have $\rm MAD_{WISE1}>0.01~mag$ (dotted line) and $\sim$2\% have $\rm MAD>0.1~mag$ (dashed line). Insert panel shows only the RSG sample with the limit of $\rm MAD_{WISE1}=0.01~mag$ in the range of $\rm 13.0<Median_{WISE1}<7.0~mag$ and $0<\rm MAD_{WISE1}<0.025~mag$. The RSG sample is separated into two subsamples. One subsample represents large variability and ``red'' targets with $\rm Median_{MAD}>0.1~mag$ and $\rm J-K_S>0.89~mag$, which may be more risky to be contaminated by AGBs (the increasing symbolic sizes indicating the increasing variabilities). The other subsample (gray) represents small variability ($\rm MAD\leq0.1~mag$) or ``blue'' ($\rm MAD>0.1~mag$ and $\rm J-K_S\leq0.89~mag$) targets. One M5e AGBs is identified (marked by the arrows) and excluded from our RSG sample, based on its significant variability with moderate magnitude. The normalized histogram (NC; upper left) indicates the range of $0<\rm MAD_{WISE1}<0.05~mag$, where the distributions of all (gray), brighter ($\rm Median_{WISE1}<9.0~mag$; red), and fainter targets ($\rm Median_{WISE1}\geq9.0~mag$; blue) are shown (for clarity, slightly different bin sizes are used for different groups). The dashed-dotted lines indicate the median values of corresponding groups. See text for details.
 \label{mag_mad}}
\end{figure*}

Figure~\ref{mag_mad} shows the median absolute deviation (MAD), which is a robust measurement of variability against outliers \citep{Rousseeuw1993}, versus median magnitude in WISE1 band (left) and the 2MASS CMD (right). From the left panel, it can be seen that the MIR variability of the RSG sample increases with luminosity. In total, $\sim$21\% of targets have $\rm MAD_{WISE1}>0.01~mag$ (dotted line), which is at least three times larger than the robust sigma of photometric error ($\sim$0.003 mag in WISE1 band) and most likely indicates real variability considering the binned data we use (also see the insert panel, which shows only the RSG sample with the limit of $\rm MAD_{WISE1}=0.01~mag$ in the range of $\rm 13.0<Median_{WISE1}<7.0~mag$ and $0<\rm MAD_{WISE1}<0.025~mag$). The rest of targets are considered as variability undetected/unreliable. Meanwhile, only a few targets ($\sim$2\%) show large variation (e.g., $\rm MAD>0.1~mag$; dashed line) with $\rm Median_{MAD}\approx0.2~mag$. For the brighter targets (e.g., $\rm Median_{WISE1}<9.0~mag$), the median MAD is about 0.03 mag, while the fainter targets (e.g., $\rm Median_{WISE1}\geq9.0~mag$) show much less variation ($\rm Median_{MAD} \approx 0.005~mag$). The normalized histogram (NC) indicates the range of $0<\rm MAD_{WISE1}<0.05~mag$, where the median values and histograms of all (gray), brighter (red), and fainter targets (purple) are shown.

Even through RSGs typically have semi-regular variabilities and moderate amplitudes, there are still some exceptions showing significant variabilities in different magnitude ranges \citep{Kiss2006, Yang2012, Soraisam2018, Ren2019}. Thus, we were only able to identify some extreme outliers based on the variabilities, luminosities, and colors at the same time. We separated the RSG sample into two subsamples. One subsample (the ``risky'' subsample) represented large variability and ``red'' targets with $\rm MAD>0.1~mag$ and $\rm J-K_S$ color redder than the median value of the RSG sample ($\rm J-K_S=0.89~mag$). It was considered as more risky to be contaminated by the AGBs, since compared to the RSGs, AGBs would have redder color and larger variability. The other subsample (the ``safe'' subsample; gray) represented small variability ($\rm MAD\leq0.1~mag$) or ``blue'' targets ($\rm MAD>0.1~mag$ and $\rm J-K_S\leq0.89~mag$). For the ``safe'' subsample, in the fainter magnitude range (e.g., $\rm WISE1\gtrsim11.0$ or $\rm K_S\gtrsim11.0~mag$), simultaneous inspection of both panels of Figure~\ref{mag_mad} indicates that the majority of targets with relatively larger variabilities are located in the blue end of the RSG sample, which is reasonable since they are likely to be the progenitors or descendants of RSGs crossing the instability strip, and may be classified as CCeps as shown in the left panel of Figure~\ref{cmd_tmass_ogle}. Meanwhile, in the brighter magnitude range (e.g., $\rm WISE1<11.0$ or $\rm K_S<11.0~mag$), a few spectroscopic RSGs also present excessive variabilities compared to the main sample of RSGs (notice that, as discussed before, the true nature of some spectroscopic RSGs are still in debate), for which other studies also indicate that the peak-to-peak amplitudes of some RSGs may reach up to four magnitudes in the optical \citep{Kiss2006, Yang2011, Yang2012, Chatys2019, Ren2019} and prevent us to further constrain the sample. However, the contamination of AGBs should be small compared to the fainter magnitudes, while as the luminosity increases, the contamination will be rapidly decreased. For the ``risky'' subsample (fourteen targets in total, six of them are spectroscopic RSGs), targets are shown in both panels of Figure~\ref{mag_mad} with increasing symbolic sizes indicating the increasing variabilities. The majority of the targets show increasing variabilities along with the increasing luminosities, except one (marked by the arrows in both panels) that shows significant variability with moderate magnitude. Further investigation indicated that it was a M5e AGB star \citep{Whitelock1989}. We excluded this target from our RSG sample. Consequently, there are 1,239 RSG candidates in the final sample as listed in the Table~\ref{fsample} (see Table 3 of \citealt{Yang2019a} regarding to the form and content of the table). 

We emphasize that there may still be a small contamination in our RSG sample from the AGBs and/or RGBs, especially in the faint and/or red end, which is inevitable and may not be disentangled solely based on the photometric data. Moreover, as mentioned before, until the next generation of large-scale spectroscopic data became available, there is no absolutely way to separate AGBs from RSGs. However, since our sample is constrained by several factors (e.g., astrometry, evolutionary model, luminosity, color, variability), we are confident that, statistically, our sample is sufficient enough to represent the entire RSGs population in the SMC up to now.

\section{Discussion}

\subsection{Mass Loss Rate, Luminosity and Variability of RSGs in the SMC}

\begin{figure*}
\center
\includegraphics[bb=55 435 555 650]{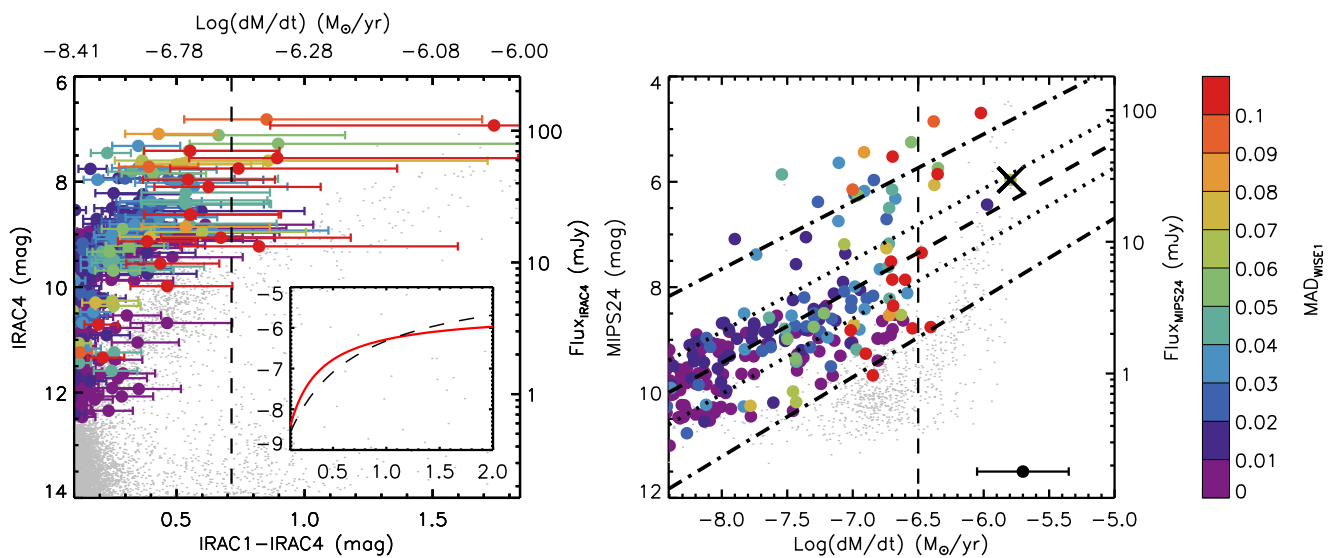}
\caption{IRAC4 versus $\rm IRAC1-IRAC4$ (left) and MIPS24 versus MLR (right) diagrams color coded with WISE1-band variability (same below). The $\rm IRAC1-IRAC4$ color is converted to the MLR by using a modified algorithm from \citet{Groenewegen2018}, where the insert panel shows the comparison between the new (red solid line) and old (black dashed line) algorithms (x-axis is the $\rm IRAC1-IRAC4$ color and y-axis is the MLR). The error bars show typical error of 0.35 dex. Most of the targets appear to have MLR below $\sim10^{-6.5}~M_\sun/yr$ (vertical dashed line; for convenience, two targets showing substantial MLR, e.g., $\rm >10^{-6.0}~M_\sun/yr$, are not shown in the IRAC4 versus $\rm IRAC1-IRAC4$ diagram). There is a linear relation (dashed line) between MIPS24 magnitude and MLR, while the dotted lines and dashed-dotted lines indicate the 1$\sigma$ and 3$\sigma$ uncertainties, respectively. Moreover, a bunch of targets lying above the upper limit of 3$\sigma$ may be related to episodic mass loss events during the RSGs phase.
 \label{mlr}}
\end{figure*}

Following \citet{Yang2018}, we analyzed the relation between MLR, luminosity, and variability of the final sample of RSGs in the SMC. Figure~\ref{mlr} shows the regimes dominated by the relatively longer wavelengths, that are IRAC4 versus $\rm IRAC1-IRAC4$ (left) and MIPS24 versus MLR (right) diagrams color coded with WISE1-band variability. The IRAC4 (8.0 $\mu m$) and MIPS24 (24 $\mu m$) bands represent the relatively hot and cold dust components, respectively. The $\rm IRAC1-IRAC4$ color was converted to the MLR by using an algorithm for O-rich supergiant stars as
\begin{equation}
Log\dot{M}=-5.598-\frac{0.817}{(IRAC1-IRAC4)+0.191},
\end{equation}
derived by \citet{Groenewegen2018}, based on a large sample of evolved stars in several Local Group galaxies with a variety of metallicities and star-formation histories. The typical uncertainty of $\rm Log\dot{M}$ is about 0.35 dex with color constraint (the algorithm may not be available for targets with $\rm IRAC1-IRAC4<0.1~mag$ due to the quick drop of MLR in the bluest color). Notice that, this algorithm is only considering O-rich RSGs and different from the original one used in the paper. This is mostly due to that the original algorithm treats AGBs and RSGs together as M-type stars, for which there is an apparent split in the MLR for AGBs and RSGs, with the RSGs usually showing higher MLRs. Moreover, when considering only RSGs, the error is also smaller ($\sim$0.35 dex versus originally $\sim$0.49 dex; Martin Groenewegen, private communication). Insert panel of Figure~\ref{mlr} shows the comparison between the new and old algorithms.

From the two diagrams, most of the targets appear to have MLR below $\sim10^{-6.5}~M_\sun/yr$ (vertical dashed lines), while targets with higher MLR are also the ones with larger variability and higher luminosity, which may be due to the contribution from both variability and luminosity \citep{Josselin2007, vanLoon2008, Yoon2010, Yang2018}. Only two targets in the diagrams show substantial MLR ($>10^{-6.0}~M_\sun/yr$). For the IRAC4 versus $\rm IRAC1-IRAC4$ diagram, there are 382 valid targets ($\sim$32\%) redder than $\rm IRAC1-IRAC4=0.1~mag$. Fifty-five of them ($\sim$14\%) are brighter than $\rm IRAC4=8.5~mag$ ($\rm Flux_{IRAC4}\gtrsim 25.84~mJy$). For the MIPS24 versus MLR diagram, there are 231 valid targets redder than $\rm IRAC1-IRAC4=0.1~mag$. Forty-one targets ($\sim$18\%) are brighter than $\rm MIPS24=8.0~mag$ ($\rm Flux_{MIPS24}\gtrsim 4.52~mJy$).

As can be seen in the MIPS24 versus MLR diagram, there is a linear relation (dashed line) between MIPS24 magnitude and MLR if we adopt the targets with $\rm IRAC1-IRAC4\geq0.1~mag$ as, 
\begin{equation}
MIPS24=(-1.37\pm0.04)\times Log\dot{M}+(-1.52\pm0.29),
\end{equation}
while the dotted lines and dashed-dotted lines indicate the 1$\sigma$ and 3$\sigma$ uncertainties (the observational error of MIPS24 band and a constant error of $\sim$0.35 dex in the MLR are both taking into account), respectively. This linear relation can be understood as the result of the degeneracy of the variability and luminosity in relation to boost the MLR as mentioned before, since the linear relation here corresponds to an exponential relation in the flux domain. In other words, the MIPS24 (24 $\mu$m) flux grows exponentially along with the increase of the brightness and variability of the star. Moreover, we also notice that there are a bunch of targets lying above the upper limit of 3$\sigma$, which may be related to episodic mass loss events during the RSGs phase \citep{Smith2009, Smith2014}, despite moderate errors they have.

Based on the derived MLR, we are able to roughly estimate the gas and dust budget produced by the RSG population in the SMC. For our sample, a total of about $\rm \sim2.2^{+2.7}_{-1.2}\times10^{-5}~M_{\odot}/yr$ is counted for 382 targets with $\rm IRAC1-IRAC4\geq 0.1$ (the rest of the 818 targets with $\rm IRAC1-IRAC4<0.1$ will contribute about $\rm \lesssim1.0\times10^{-7}~M_{\odot}/yr$, if we adopt $\rm \sim1.0\times10^{-10}~M_{\odot}/yr$ as the average MLR, due to the quick drop of MLR in the bluest color; \citealt{Groenewegen2018}). Among them, about 12\% of the MLR ($\rm \sim2.7^{+3.3}_{-1.5}\times10^{-6}~M_{\odot}/yr$) originate from two extremely dusty RSG candidates with $\rm MLR>10^{-6}~M_\sun/yr$. Moreover, considering the saturation problems in the IRAC4 band (two targets with $\rm J-K_S>2.0$ are saturated in IRAC4 band) and the incompleteness of our sample, the total MLR of RSG population could be ever higher. Even for the most conservative case where the derived MLR is overestimated by one order of magnitude and not considering the extreme dusty RSG candidates, the total MLR of the RSG population would still be around $\rm \sim1.9^{+2.4}_{-1.1}\times10^{-6}~M_{\odot}/yr$. Since the gas-to-dust ratio (GDR) in \citet{Groenewegen2018} is taken as a constant value of 200, this MLR can be transfer to a dust-production rate of $\rm \sim1.0^{+1.2}_{-0.5}\times10^{-8}~M_{\odot}/yr$, which is $\sim$30\% of \citet{Boyer2012} and $\sim$20\% of \citet{Srinivasan2016}. However, considering the large uncertainty of GDR, the results still can be considered as comparable, and the discrepancies could be attributed to differences in sample sizes, adopted optical constants, and details in the radiative transfer modeling (\citealt{Groenewegen2018}; also see their Section 6.1). We also need to emphasize that the sample of \citet{Groenewegen2018} used to derive the relation between $\rm IRAC1-IRAC4$ color and MLR only contains very few RSGs with $\rm IRAC1-IRAC4>1.5~mag$, indicating that the MLR of extreme red objects (e.g., the obscured targets) may be underestimated. Moreover, in fact, instead of the overestimation as we presumed in the most conservative case, the $\rm IRAC1-IRAC4$ color may underestimate the MLR of the brightest RSGs as mentioned in Section 3.2 (the small difference between $\rm J-IRAC4$ and $\rm IRAC1-IRAC4$ colors can be ignored), so that the total MLR would be even higher. The detailed analysis of MLR of RSGs in the SMC will be presented in a future paper.

\subsection{Geneva evolutionary model of RSGs at the SMC metallicity}

\begin{figure*}
\center
\includegraphics[bb=130 370 460 690, scale=0.65]{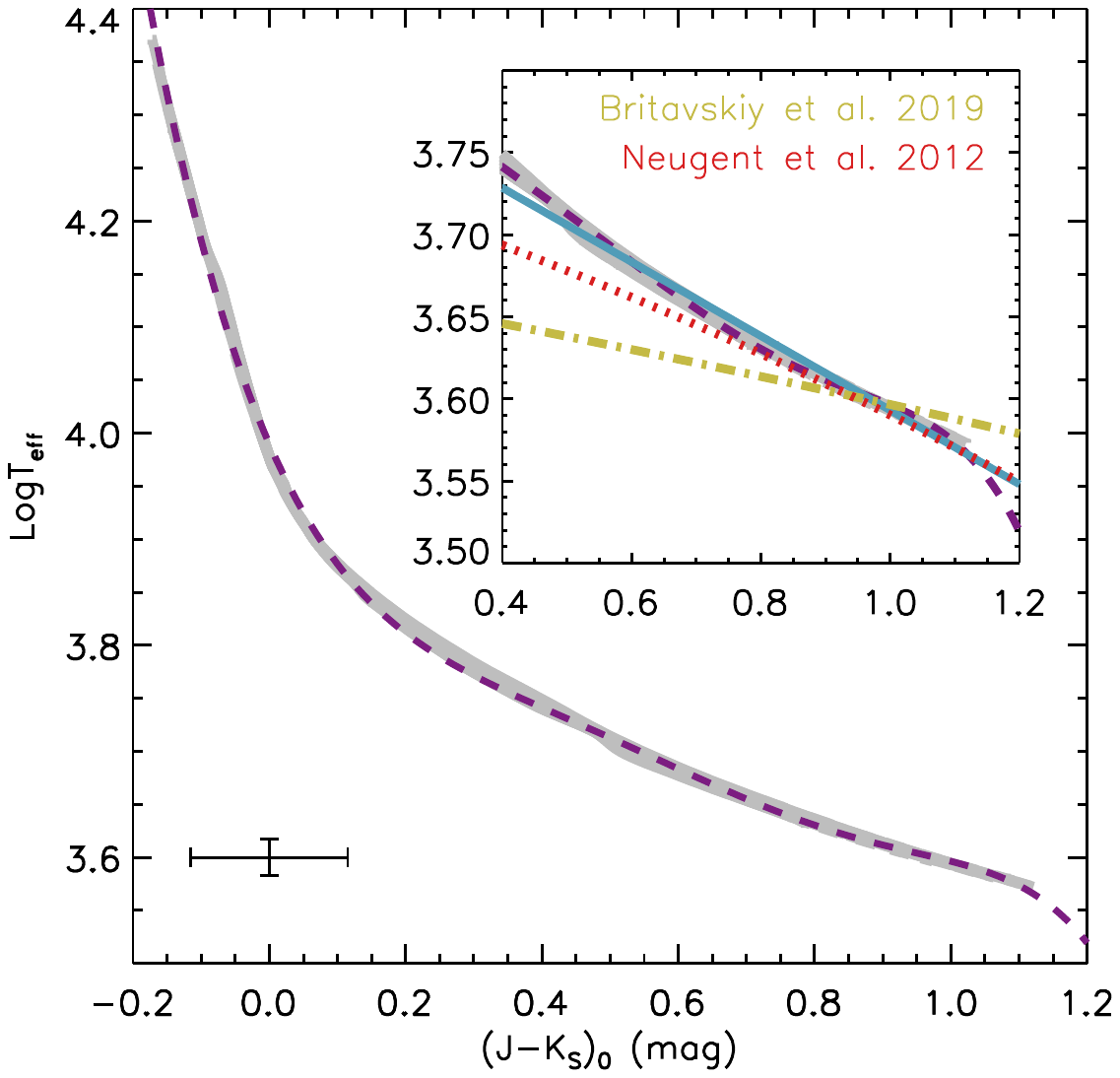}
\includegraphics[bb=130 370 525 690, scale=0.65]{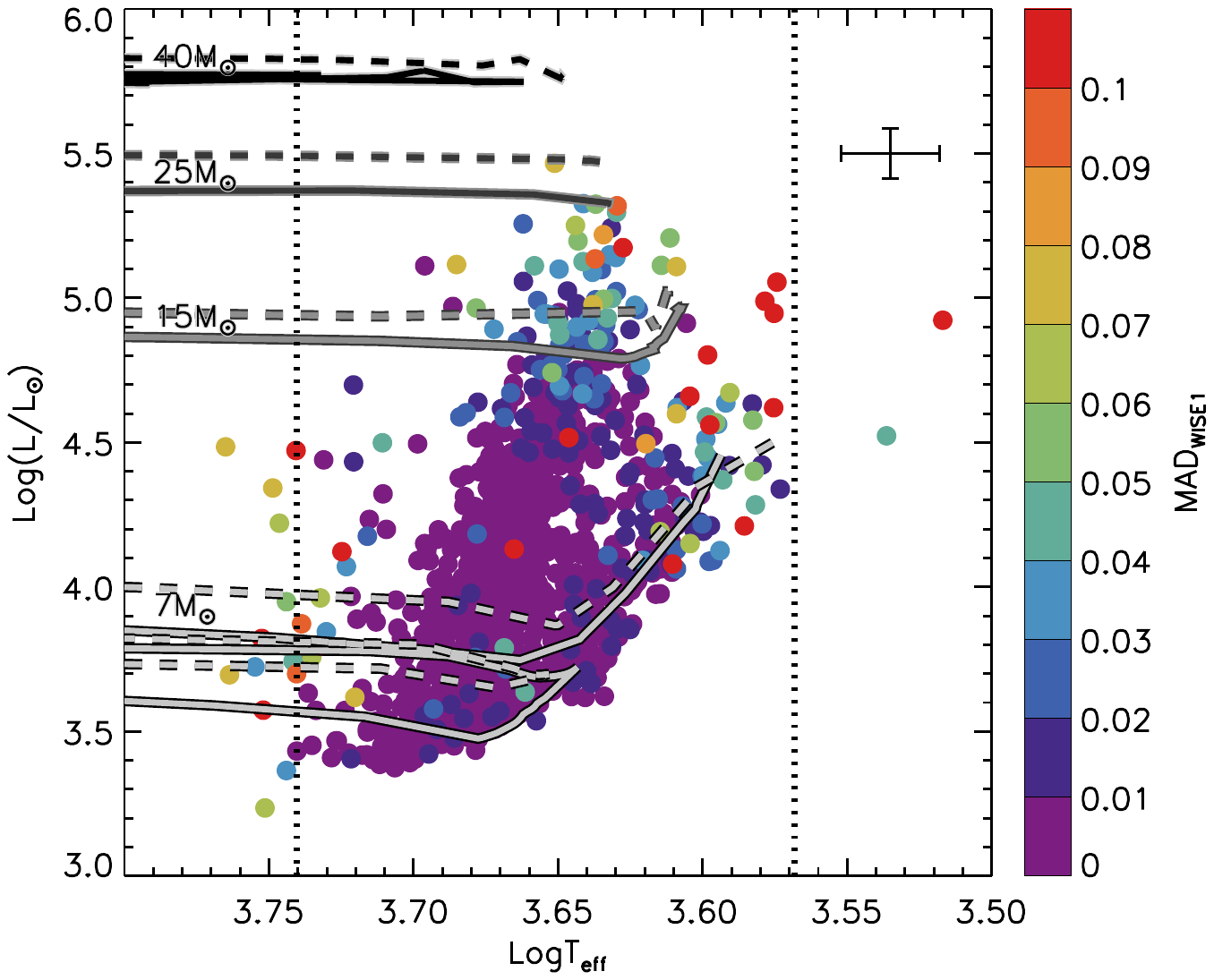}
\caption{Left: $T_{\rm eff}$ versus reddening-free $\rm J-K_S$ color ($\rm (J-K_S)_0$) over the range of $3.57\leq LogT_{\rm eff}\leq 4.35$ and $\rm -0.16\leq(J-K_S)_0\leq 1.13$ derived from the MIST models. A six-order polynomial fitting is shown as dashed line, which works very well, except in the very red end. The insert panel shows the range of $3.5\leq LogT_{\rm eff}\leq 3.8$ and $\rm 0.4\leq(J-K_S)_0\leq 1.2$, where a linear fitting is adopted shown as solid line instead of the polynomial fitting. Results from \citet{Britavskiy2019} (dashed-dotted line) and \citet{Neugent2012} (dotted line) are also shown in the diagram. Our result is almost identical to \citet{Neugent2012} in the range of $\rm 0.8\leq(J-K_S)_0\leq1.2$. Right: RSG sample (color coded with WISE1-band variability) overlapped with color coded non-rotation (solid lines) and rotation ($V/V_C=0.40$; dashed lines) Geneva evolutionary tracks of 7 to 40 $M_\sun$ at $\rm Z=0.002$. Vertical dotted lines indicate the $T_{\rm eff}$ range of 3,700$\sim$5700 K for the RSG sample. The vast majority of targets selected by MIST model are following the Geneva tracks, and a few outliers can be explained by the combination of variability and reddening. Error bars show the typical errors of $\sim$0.017 dex in $\rm Log \textit{T}_{\rm eff}$ and $\sim$0.086 dex in $\rm Log(L/L_{\sun})$.
 \label{geneva}}
\end{figure*}

We also compared our final RSG sample with the Geneva stellar evolutionary tracks at the metallicity of $\rm Z=0.002$ \citep{Georgy2013}. The first step was to convert the observed parameters into physical quantities, namely from the magnitudes and colors to the luminosity and $T_{\rm eff}$. We adopted a constant bolometric correction of $\rm BC_{K_S}=2.69\pm0.06$ \citep{Davies2013} to convert $\rm K_S$-band magnitude into luminosity, due to the fact that $\rm BC_{K_S}$ of RSGs is independent of IR color with small uncertainty (about $15\sim25\%$; \citealt{Buchanan2006, Davies2013}). There may be a concern that this $\rm BC_{K_S}$ is derived from RSGs with $\rm Log(L/L_{\sun})\geq4.6$, and here is extrapolated about one order of magnitude down. However, as discussed in \citet{Davies2010}, throughout the mass range of RSGs, the pressure scale height remains largely unchanged. Moreover, we compared it before with another BC derived by \citet{Neugent2012}, which gave more or less the same result within the error range down to $\rm Log(L/L_{\sun})\approx4.0$, except in the very faint red end (see also Figure 19 of \citealt{Yang2018}). Meanwhile, in our sample, we only selected targets with Ranks of -1 to 3, excluding most of the ambiguous targets in the faint red end. In that sense, the result can be considered as acceptable. The conversion between color and $T_{\rm eff}$ is more difficult, since the most of established relations are used for higher metallicity environments and also derived based on relatively small sample size \citep{Levesque2006, Neugent2012, Tabernero2018, Britavskiy2019}. Thus, we derived the relation between color and $T_{\rm eff}$ by taking advantage of the MIST evolutionary tracks and synthetic photometry. The left panel of Figure~\ref{geneva} shows the $T_{\rm eff}$ versus reddening-free $\rm J-K_S$ color ($\rm (J-K_S)_0$) over the range of $3.57\leq LogT_{\rm eff}\leq 4.35$ and $\rm -0.16\leq(J-K_S)_0\leq 1.13$ derived from the MIST model. We fitted a six-order polynomial shown as the blue dashed line to the data as,
\begin{equation}
\begin{split}
LogT_{eff}=&-0.71(J-K_S)_0^6+0.41(J-K_S)_0^5+3.30(J-K_S)_0^4 \\
&-5.85(J-K_S)_0^3+3.92(J-K_S)_0^2-1.47(J-K_S)_0 \\
&+3.99,
\end{split}
\end{equation}
which worked very well, except in the very red end. The insert panel shows the range of $3.5\leq LogT_{\rm eff}\leq 3.8$ and $\rm 0.4\leq(J-K_S)_0\leq 1.2$ where the majority of RSGs are supposed to be located. It can be seen that the relation is almost linear within the range and the polynomial fitting largely underestimate the $T_{\rm eff}$ in the very red end (e.g., $\rm (J-K_S)_0\approx1.2$). Alternatively, a linear relation shown as the blue solid line was fitted for this range as,
\begin{equation}
LogT_{eff}=-0.23(J-K_S)_0+3.82,
\end{equation}
and adopted to convert the observed $\rm J-K_S$ color to $T_{\rm eff}$ with reddening correction. Meanwhile, the results from \citet{Britavskiy2019} and \citet{Neugent2012} are also shown in the diagram. Our result is almost identical to \citet{Neugent2012} in the range of $\rm 0.8\leq(J-K_S)_0\leq1.2$, where their sample is located, proving that our fitting is appropriate. The difference between the studies is likely due to different metallicities, sample size, and the treatments of parameters in the models.

The typical uncertainty in $\rm J-K_S$ is $\sim$0.033 mag. In combination with uncertainties from the reddening (e.g., 50\% uncertainty of $\rm E(J-K_S)$ as $\sim$0.11 mag) and variability (assuming that 2MASS bands have typical variability of $\sim$0.01 mag that is similar to the WISE1 band), this propagates to error of $\sim$0.017 dex in $\rm Log \textit{T}_{\rm eff}$. For $\rm Log(L/L_{\sun})$, an error of $\sim$0.086 dex is originated from the uncertainty of $\sim$0.06 mag in $\rm BC_{K_S}$, the typical uncertainty of $\sim$0.022 mag in $\rm K_S$, $\sim$0.01 mag of variability, $\sim$0.05 mag for 50\% uncertainty of extinction, and $\sim$0.2 mag for 50\% uncertainty of the depth in 3D structure of the SMC. The dominant factors of errors for $\rm Log \textit{T}_{\rm eff}$ and $\rm Log(L/L_{\sun})$ come from the reddening and 3D structure of the SMC, respectively. More accurate correction for each individual target cannot be done without precise measurements of the distance and local extinction/reddening. The corresponding errors are shown in both panels of the Figure~\ref{geneva}.

The right panel of Figure~\ref{geneva} shows our RSG sample (color coded with WISE1-band variability) overlapped with color coded non-rotation (solid lines) and rotation ($\rm V/V_C=0.40$; dashed lines) Geneva evolutionary tracks of 7 to 40 $\rm M_\sun$ at $\rm Z=0.002$. We adopted a color excess of $\rm E(J-K_S)=0.212$ ($\rm A_{K_S}=0.1~mag$, $\rm A_J/A_{K_S}=3.12$ and $\rm A_{K_S}/A_V\approx0.1$) for de-reddening \citep{Wang2019}. We need to emphasize that for the reddest RSG candidates which may be very close to the star formation region, or largely self-obscured, the reddening can be much larger. From the diagram, it can be seen that the vast majority of targets selected by MIST model are following the Geneva tracks, proving that our conversion is appropriate. The $T_{\rm eff}$ range of the RSG sample spanning around 3,700$\sim$5700 K (dotted lines). We notice that at the bright end, there is only one target above 25 $M_\sun$ track. However, this also could be due to the large extinction and/or variability of the obscured RSGs, like the outliers on the upper right region of the diagram. At the faint end, by using BC from \citet{Davies2013} other than the MIST BC table, targets still reach down to the 7 $M_\sun$ track within the error range, indicating a lower initial mass limit of $\sim$7 $M_\sun$ for the RSG population as found in \citet{Yang2019a}. Meanwhile, a few outliers deviated largely from the tracks can be explained by the combination of reddening and variability as mentioned in \citet{Yang2018}.

\subsection{Comparison of RSGs populations between the SMC and LMC}

To better understand the relation between metallicity and other physical properties of RSGs, we compared the RSG sample in the LMC from \citet{Yang2018} and the final RSG sample in the SMC. It must to be emphasized that \citet{Yang2018} was published before \textit{Gaia} DR2 released, for which we were only able to retrieve a small sample of the RSGs in the LMC confirmed either by spectroscopy or photometry from previous studies. After \textit{Gaia} DR2 was released, a quick calculation by using the same method as \citet{Yang2019a} indicated that, the ratio of RSG sample from \citet{Yang2018} to the entire RSGs population in the LMC was higher than we thought (about 20$\sim$25\% instead of 10$\sim$15\%; mainly due to the reduced foreground contamination), but was still not very representative. Thus, the comparison between the SMC and LMC presented here only indicates the general difference between these two galaxies.

\begin{figure*}
\center
\includegraphics[bb=55 370 560 720]{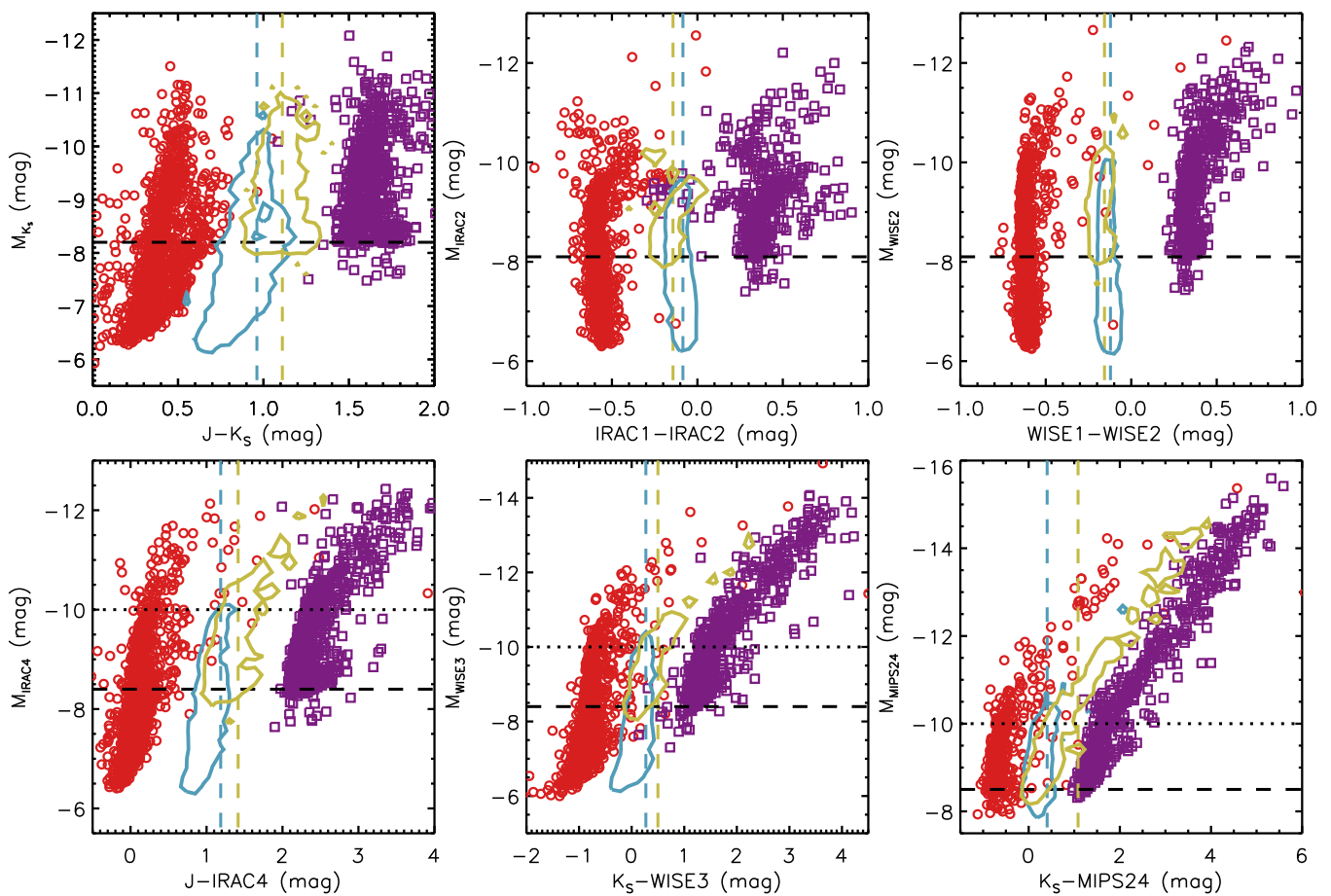}
\caption{CMDs of $\rm M_{K_S}$ versus $\rm J-K_S$ (upper left), $\rm M_{IRAC2}$ versus $\rm IRAC1-IRAC2$ (upper middle), $\rm M_{WISE2}$ versus $\rm WISE1-WISE2$ (upper right), $\rm M_{IRAC4}$ versus $\rm J-IRAC4$ (bottom left), $\rm M_{WISE3}$ versus $\rm K_S-WISE3$ (bottom middle), and $\rm M_{MIPS24}$ versus $\rm K_S-MIPS24$ (bottom right). Targets with upper limit are not shown in the diagrams (same below). Red open circles and purple open squares indicate the targets from the SMC and the LMC samples, while the blue and yellow contours indicate 10\% number density of each sample (same below), respectively. There are offsets of $\rm -0.5~mag$ (upper row) and $\rm -1.0~mag$ (bottom row) for the color indexes of the SMC sample, and offsets of $\rm +0.5~mag$ and $\rm +1.0~mag$ for the LMC sample, respectively (contours are not offset; same below). The vertical dashed lines indicate the median values of each color index (blue for the SMC and yellow for the LMC) with the same limiting magnitudes in both the SMC and the LMC (horizontal dashed lines). The LMC sample is not magnitude-completed and lack of a large number of fainter and bluer targets. The differences of color indexes between the SMC and the LMC are listed in Table~\ref{diff_smc_lmc}. Even though the LMC sample is less complete, it still contains more dusty targets ($<-10.0~mag$; horizontal dotted lines) than the SMC sample at the longer wavelengths.
 \label{cmd_smc_lmc}}
\end{figure*}

Figure~\ref{cmd_smc_lmc} shows multiple CMDs from NIR to MIR bands (there is no optical data from \citealt{Yang2018}). A distance modulus of $18.493\pm0.055$ for the LMC is adopted \citep{Pietrzynski2013}. For convenience, there are offsets of -0.5 mag in the upper row and -1.0 mag in the bottom row for the color indexes of the SMC sample, while offsets of +0.5 mag and +1.0 mag for the LMC sample, respectively (contours are not offset). It can be seen that, the LMC sample is not magnitude-completed and lack of a large number of fainter and bluer targets. However, considering the upper parts of both the SMC and the LMC sample with comparable magnitudes, there is still a visible difference. Thus, in order to minimize the influence of incompleteness of the LMC sample and not exaggerate the effect of metallicity, we compared the median values of each color index (vertical dashed lines; blue for the SMC and yellow for the LMC) with the same limiting magnitude in both the SMC and the LMC (horizontal dashed lines) as listed in Table~\ref{diff_smc_lmc}. For $\rm M_{K_S}$ versus $\rm J-K_S$ diagram where the stellar photospheric emission remains dominating, the SMC sample shows bluer color than the LMC sample due to the shifting of average spectral type of RSGs towards earlier types at lower metallicities \citep{Elias1985, Massey2003, Levesque2006, Levesque2012, Dorda2016}. For $\rm M_{IRAC2}$ versus $\rm IRAC1-IRAC2$ and $\rm M_{WISE2}$ versus $\rm WISE1-WISE2$ diagrams, the differences between the SMC and LMC are very small. This may indicate that CO absorption around 4.6 $\mu$m~is much less affected by the metallicity. Finally, for the longer wavelengths (e.g., IRAC4, WISE12, and MIPS24 bands) where the dust emission dominates, the differences are more obvious. The LMC sample contains more dusty targets (e.g., $\rm <-10.0~mag$mag as shown by the horizontal dotted lines) than the SMC ($\rm N_{IRAC4_{SMC}}/N_{IRAC4_{LMC}}\approx0.30$, $\rm N_{WISE3_{SMC}}/N_{WISE3_{LMC}}\approx0.35$ and $\rm N_{MIPS24_{SMC}}/N_{MIPS24_{LMC}}\approx0.38$), despite the fact that the LMC sample is more incomplete. This is likely indicating a positive relation between MLR and metallicity \citep{vanLoon2005, Mauron2011}.

\begin{figure*}
\center
\includegraphics[bb=55 370 560 720]{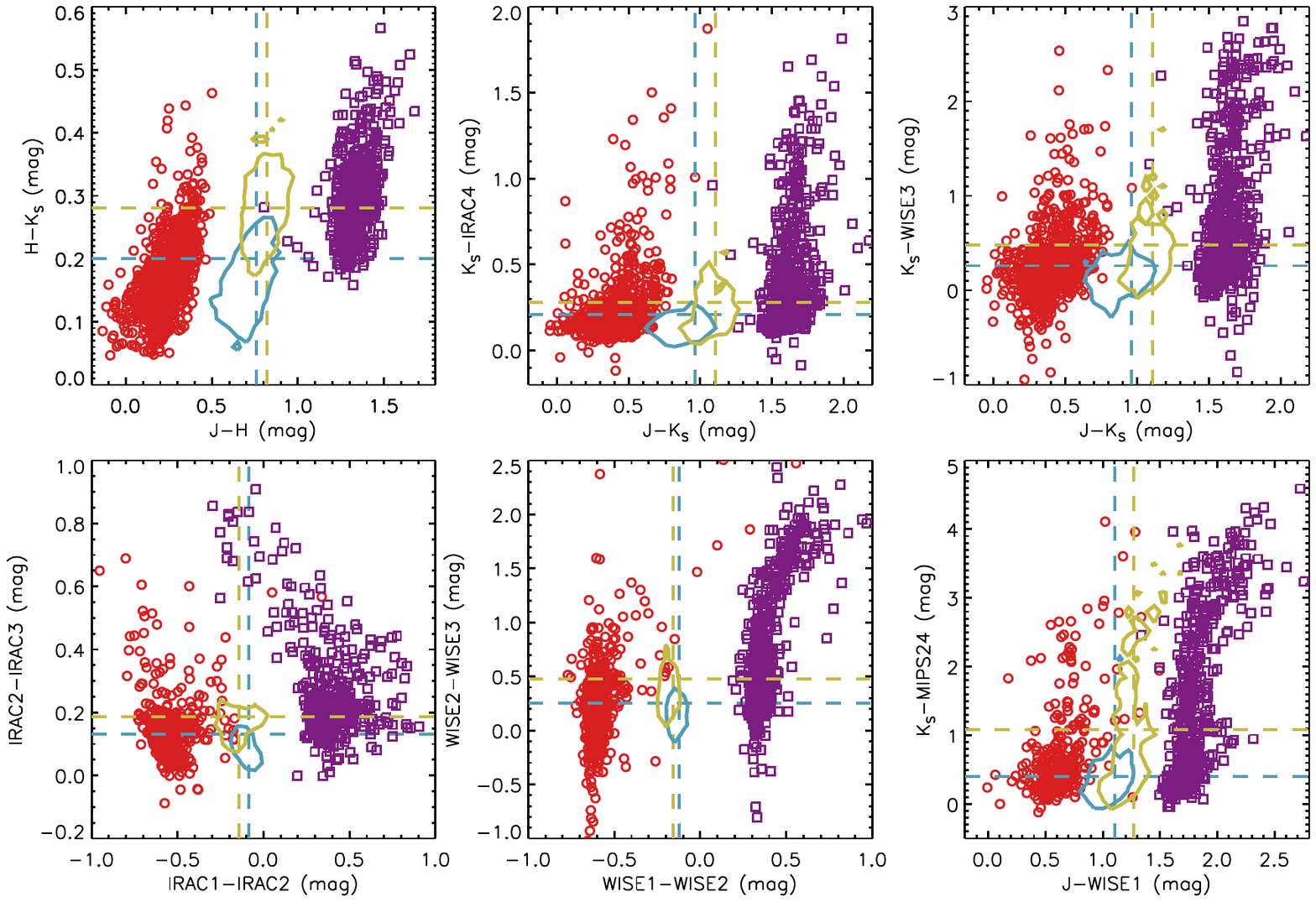}
\caption{CCDs of $\rm J-H$ versus $\rm H-K_S$ (upper left), $\rm J-K_S$ versus $\rm K_S-IRAC4$ (upper middle), $\rm J-K_S$ versus $\rm K_S-WISE3$ (upper right), $\rm IRAC1-IRAC2$ versus $\rm IRAC2-IRAC3$ (bottom left), $\rm WISE1-WISE2$ versus $\rm WISE2-WISE3$ (bottom middle), and $\rm J-WISE1$ versus $\rm K_S-MIPS24$ (bottom right). The differences of additional color indexes are also listed in Table~\ref{diff_smc_lmc}.
 \label{ccd_smc_lmc}}
\end{figure*}

Figure~\ref{ccd_smc_lmc} shows multiple CCDs. Similar to the CMDs, the LMC sample always shows redder color than the SMC sample, except for the $\rm IRAC1-IRAC2$ and $\rm WISE1-WISE2$ colors. The differences of additional color indexes are also listed in Table~\ref{diff_smc_lmc} with corresponding limiting magnitudes.

\begin{figure*}
\center
\includegraphics[bb=55 405 540 645, scale=0.8]{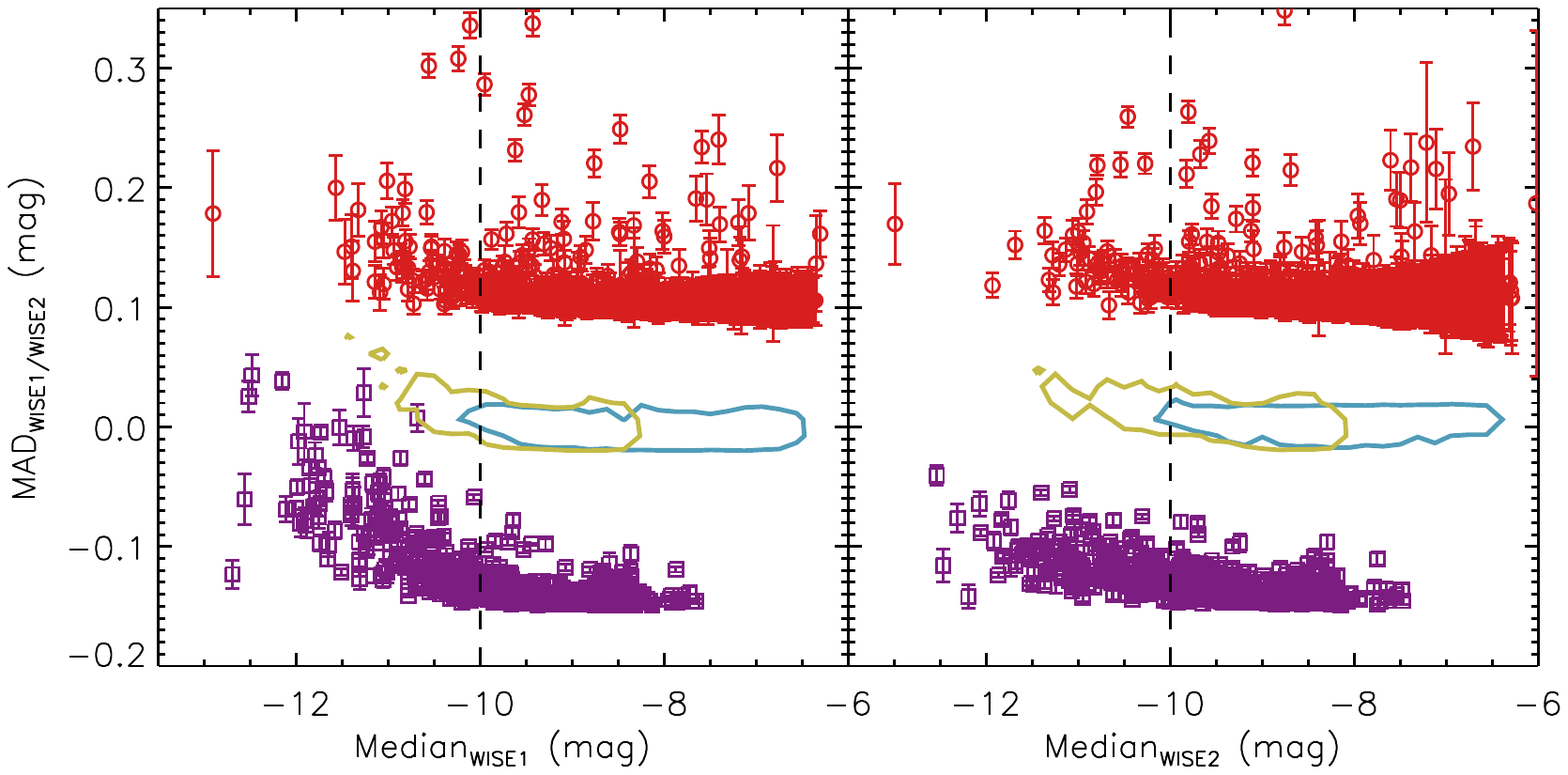}
\caption{Median magnitude versus MAD in both WISE1 (left) and WISE2 (right) bands. For WISE1 band, the differences of median values of MAD between SMC and LMC are about -0.0050 (brighter sources; $\rm \leq-10.0~mag$ shown as the vertical dashed line), -0.0012 mag (fainter sources; $\rm >-10.0~mag$) and -0.0045 mag (all sources). For WISE2 band, the differences are about 0.0012 (brighter sources), -0.0008 (fainter sources) and -0.0035 mag (all sources), respectively. 
 \label{var_smc_lmc}}
\end{figure*}

Finally, Figure~\ref{var_smc_lmc} shows the median magnitude versus MAD in both WISE1 (left) and WISE2 (right) bands. In general, the difference of median values of MAD between SMC and LMC is more significant for the brighter sources (e.g., $\rm \leq-10.0~mag$, vertical dashed line, where the bright targets start to show notable variabilities as visually inspected) as $\rm MAD_{SMC}-MAD_{LMC}\approx-0.0050~mag$, than the fainter sources (e.g., $\rm >-10.0~mag$) as $\rm MAD_{SMC}-MAD_{LMC}\approx-0.0012~mag$ for WISE1 band, while the difference between the whole samples is $\rm \sim-0.0045~mag$, indicating that the variability of RSGs is likely to be also metallicity-dependent. Meanwhile, the differences of WISE2 band are smaller as $\rm \sim0.0012~mag$ (brighter sources), $\rm \sim-0.0008~mag$ (fainter sources) and $\rm \sim-0.0035~mag$ (all sources), respectively, due to the lower sensitivity/larger scattering in WISE2 band and the larger distance of the SMC.

\section{Summary}

We present the most comprehensive RSG sample in the SMC up to now. The sample is mainly built based on a source catalog for the SMC \citep{Yang2019a} with conservative ranking (targets were identified as RSGs in at least two CMDs). Moreover, additional optical and MIR spectroscopic RSGs are retrieved from Simbad and data taken by \textit{Spitzer}/IRS, as well as RSG candidates selected based on the visual inspection of both \textit{Gaia} and 2MASS CMDs. In total, the final sample contains 1,239 RSG candidates with 327 unique spectroscopic RSGs. We estimate that there are $\sim$ 1,800 or more RSGs in the SMC.

We have studied the NIR to MIR infrared properties of the RSG sample. The investigation of 2MASS CMD indicates that, at the bright magnitude end, the spectroscopic RSGs only represent about one-fourth of the whole RSG population, but follow almost exactly the MIST tracks at the red end. It also shows that many O-AGBs defined by the theoretical color cuts but located inside the MIST model region (47 out of 206 targets; $\sim$23\%) are identified as spectroscopic RSGs by previous studies, as well as 24 C-OSARGs, 7 O-SRVs, and 3 O-Miras defined by OGLE. However, since there is a continuum with similarity and overlapping between RSGs and AGBs in both spectra and LCs, the clear separation between them is still a pending issue. Two targets that are fainter than both the $\rm K_S$-TRGB and IRAC1-TRGB are excluded from the RSG sample. We also identify one heavily obscured target by adopting $\rm K_S>12.7~mag$ or $\rm IRAC1>12.6~mag$, $\rm J-IRAC4\geq3.0~mag$, and $\rm IRAC4<10.0~mag$, based on the fact that obscured objects will be brighter and redder at the longer wavelengths compared to the shorter wavelengths. 

Further analysis of CMDs with longer wavelengths indicates the complex behavior of RSG sample around 3-4$\mu$m, the growth of MLR and circumstellar envelope with increasing luminosity, and a similar large MLR between RSGs and x-AGBs for the brightest RSG candidates. The inconsistency between infrared excesses of the relatively shorter (8 $\mu$m) and longer (12 and 22 $\mu$m) wavelengths suggests that 8 $\mu$m may not be enough to characterize the MLR of brightest RSGs. The investigation of CCDs shows that there are much fewer RSGs candidates (only about 4\%) showing PAH emission features compared to the Milky Way and LMC ($\sim$15\%), which may be due to strong metallicity dependence of PAH abundance. We analyze the MIR time-series data from WISE for our RSG sample. The MIR variability of the RSG sample is found to increase with luminosity, while about 21\% of targets show reliable variability ($\rm MAD_{WISE1}>0.01~mag$). We separate the RSG sample into two subsamples (``risky'' and ``safe'') and identify one M5e AGB star in the ``risky'' subsample, based on the variabilities, luminosities and colors at the same time. 

For the final RSG sample, the degeneracy of MLR, variability and luminosity is presented. We found that most of the large variability targets are also the bright ones with large MLR. Although there is a linear relation between MIPS24 magnitude and MLR for the RSG sample, the fluxes of some targets is above the upper limit of 3$\sigma$ of the linear relation, which may be related to the episodic mass loss events during the RSGs phase. We also roughly estimate the total dust budget produced by the entire RSG population as $\rm \sim1.9^{+2.4}_{-1.1}\times10^{-6}~M_{\odot}/yr$ in the most conservative case, based on the derived MLR from $\rm IRAC1-IRAC4$ color, which is comparable with previous works and will be further investigated in a future paper.

Based on the MIST evolutionary tracks and synthetic photometry, we derive a linear relation between $T_{\rm eff}$ and observed $\rm J-K_S$ color with reddening correction for the RSG sample. By using a constant bolometric correction and this relation, the Geneva evolutionary model is compared with our RSG sample, showing that there is a lower initial mass limit of $\sim$7 $M_\sun$ for the RSG population as found in \citet{Yang2019a}. Moreover, the vast majority of targets are following the tracks, and few outliers can be explained by the combination of variability and reddening as mentioned in \citet{Yang2018}.

Finally, we compare the RSG sample in the LMC \citep{Yang2018} and the final RSG sample in the SMC. Despite the incompleteness of LMC sample in the faint end, the result indicates that the LMC sample always shows redder color than the SMC sample, except for the $\rm IRAC1-IRAC2$ and $\rm WISE1-WISE2$ colors (indicating CO absorption around 4.6 $\mu$m~may be much less affected by the metallicity). The difference is more obvious at the longer wavelengths, which is likely indicate a positive relation between MLR and metallicity. Moreover, the similar result is found for the MIR variability, as LMC sample showing larger variability than the SMC sample, and it is more significant for the brighter sources than the fainter sources, which indicates that the variability of RSGs also likely to be metallicity dependent.

We are currently in the progress of investigating also the evolved massive star populations (including a much more complete sample of RSGs) in the LMC. We also plan to further explore the RSG population in the nearby low-metallicity dwarf galaxies, by utilizing deep optical, NIR, and MIR photometric data. Our studies will help to provide a more comprehensive view of evolution and dust production of evolved massive stars at low metallicity environment.

\section{Acknowledgments}

We would like to thank the anonymous referee for many constructive comments and suggestions. This study has received funding from the European Research Council (ERC) under the European Union's Horizon 2020 research and innovation programme (grant agreement number 772086). B.W.J and J.G. gratefully acknowledge support from the National Natural Science Foundation of China (Grant No.11533002 and U1631104). We thank M. A. T. Groenewegen and G. C. Sloan help on the calculation of MLR. We thank Man I Lam and Stephen A. S. de Wit for helpful comments and suggestions.

This publication makes use of data products from the Two Micron All Sky Survey, which is a joint project of the University of Massachusetts and the Infrared Processing and Analysis Center/California Institute of Technology, funded by the National Aeronautics and Space Administration and the National Science Foundation. This work is based in part on observations made with the Spitzer Space Telescope, which is operated by the Jet Propulsion Laboratory, California Institute of Technology under a contract with NASA. This publication makes use of data products from the Wide-field Infrared Survey Explorer, which is a joint project of the University of California, Los Angeles, and the Jet Propulsion Laboratory/California Institute of Technology. It is funded by the National Aeronautics and Space Administration. This publication makes use of data products from the Near-Earth Object Wide-field Infrared Survey Explorer (NEOWISE), which is a project of the Jet Propulsion Laboratory/California Institute of Technology. NEOWISE is funded by the National Aeronautics and Space Administration. This research has made use of the NASA/IPAC Infrared Science Archive, which is operated by the Jet Propulsion Laboratory, California Institute of Technology, under contract with the National Aeronautics and Space Administration.

This work has made use of data from the European Space Agency (ESA) mission {\it Gaia} (\url{https://www.cosmos.esa.int/gaia}), processed by the {\it Gaia} Data Processing and Analysis Consortium (DPAC, \url{https://www.cosmos.esa.int/web/gaia/dpac/consortium}). Funding for the DPAC has been provided by national institutions, in particular the institutions participating in the {\it Gaia} Multilateral Agreement.

This research has made use of the SIMBAD database and VizieR catalog access tool, operated at CDS, Strasbourg, France, and the Tool for OPerations on Catalogues And Tables (TOPCAT; \citealt{Taylor2005}).

\begin{table*}
\caption{Numbers of Classified Variable Stars from OGLE}
\label{var_ogle}
\begin{tabular}{cccc}
\toprule\toprule
Class & Subclass & Number & Source \\
\midrule
Long Period Variables (LPVs) &  & 8,982 & O$^3$CVS\tablefootmark{a} \\
\midrule
 & Carbon-rich OGLE Small Amplitude Red Giants (C-OSARGs) & 439 &  \\
 & Oxygen-rich OSARGs (O-OSARGs) & 7,417 &  \\
 & Carbon-rich Semi-Regular Variables (C-SRVs) & 644 & \\
 & Oxygen-rich SRVs (O-SRVs) & 371 & \\
 & Carbon-rich Miras (C-Miras) & 92 & \\
 & Oxygen-rich Miras (O-Miras) & 19 &  \\
\midrule 
Classical Cepheids (CCeps) &  &  482  &  O$^4$CVS\tablefootmark{b} \\
Type \uppercase\expandafter{\romannumeral2} Cepheids (T2Ceps) &  &  11  &  O$^4$CVS \\
Eclipsing Binaries (Ecls) &  &  87  &  O$^4$CVS \\
\midrule
\end{tabular}
\tablefoot{
\tablefoottext{a}{OGLE-\uppercase\expandafter{\romannumeral3} Catalog of Variable Stars.}
\tablefoottext{b}{OGLE-\uppercase\expandafter{\romannumeral4} Catalog of Variable Stars.}
}
\end{table*}

\begin{table*}
\caption{Final sample of 1,239 red supergiant star candidates in the SMC} 
\label{fsample}
\centering
\begin{tabular}{cccccccc}
\toprule\toprule
ID & R.A.(J2000) & Decl.(J2000) & 2MASS\_J & e\_2MASS\_J & ...... & OGLE\_Ecl\_DS & Rank \\
   & (deg)       & (deg)        & (mag)    & (mag)       & ...... & (mag)         &      \\
\midrule 
 124 &  4.368634  &   -73.428555 &   12.497 &  0.022 &  ......  &    &  2  \\
 236 &  4.921237  &   -73.353052 &   11.675 &  0.022 &  ......  &    &  4  \\
 240 &  4.952093  &   -73.588549 &   12.672 &  0.024 &  ......  &    &  2  \\
 635 &  5.800023  &   -72.390737 &   12.920 &  0.029 &  ......  &    &  3  \\
 747 &  5.977038  &   -73.639891 &   12.253 &  0.023 &  ......  &    &  2  \\
...  &   ...    &  ...         &  ...     &  ...    &  ......  &  ...     &  ...  \\
\midrule 
\end{tabular}
\tablefoot{
This table is available in its entirety in CDS. A portion is shown here for guidance regarding its form and content.\\
}
\end{table*}

\begin{table*}
\caption{Differences of Color Indexes between the SMC and the LMC samples} 
\label{diff_smc_lmc}
\centering
\begin{tabular}{cc}
\toprule\toprule
Limiting Magnitude & $\rm \Delta(SMC-LMC)$ \\
\midrule 
$\rm M_{H}=-8.0~mag$ & $\rm \Delta(J-H)\approx-0.059~mag$ \\
$\rm M_{H}=-8.0~mag$ & $\rm \Delta(H-K_S)\approx-0.081~mag$ \\
$\rm M_{K_S}=-8.2~mag$ & $\rm \Delta(J-K_S)\approx-0.15~mag$ \\
$\rm M_{K_S}=-8.2~mag$ & $\rm \Delta(K_S-IRAC4)\approx-0.067~mag$ \\
$\rm M_{WISE1}=-8.3~mag$ & $\rm \Delta(J-WISE1)\approx-0.16~mag$ \\
$\rm M_{IRAC2}=-8.1~mag$ & $\rm \Delta(IRAC1-IRAC2)\approx0.058~mag$ \\
$\rm M_{IRAC2}=-8.1~mag$ & $\rm \Delta(IRAC2-IRAC3)\approx-0.055~mag$ \\
$\rm M_{WISE2}=-8.1~mag$ & $\rm \Delta(WISE1-WISE2)\approx0.035~mag$ \\
$\rm M_{WISE2}=-8.1~mag$ & $\rm \Delta(WISE2-WISE3)\approx-0.022~mag$ \\
$\rm M_{IRAC4}=-8.4~mag$ & $\rm \Delta(J-IRAC4)\approx-0.23~mag$ \\
$\rm M_{WISE3}=-8.4~mag$ & $\rm \Delta(K_S-WISE3)\approx-0.23~mag$ \\
$\rm M_{MIPS24}=-8.5~mag$ & $\rm \Delta(K_S-MIPS24)\approx-0.68~mag$ \\
\midrule 
\end{tabular}
\end{table*}

\clearpage

\end{CJK*}

\end{document}